\documentclass[journal,twoside,web]{ieeecolor}
\usepackage{generic}
\usepackage{amsmath,amssymb,amsfonts}
\usepackage{algorithmic}
\usepackage{graphicx}
\usepackage{textcomp}

\usepackage{algorithmic}
\usepackage{algorithm}
\usepackage{array}
\usepackage[caption=false,font=normalsize,labelfont=sf,textfont=sf]{subfig}
\usepackage{textcomp}
\usepackage{url}
\usepackage{verbatim}
\usepackage{graphicx}
\usepackage{lipsum}
\usepackage{csvsimple}
\usepackage{csquotes}
\usepackage[backend=bibtex,style=numeric,url=false,giveninits=true, maxnames=1]{biblatex}
\usepackage{multirow}
\usepackage{placeins}
\usepackage{booktabs}
\renewcommand{\arraystretch}{1.25}
\newcolumntype{P}[1]{>{\centering\arraybackslash}p{#1}}
\newcolumntype{M}[1]{>{\centering\arraybackslash}m{#1}}
\usepackage{caption}
\usepackage{listings}
\usepackage[table,usenames,dvipsnames]{xcolor}
\usepackage{parskip}
\usepackage{soul}
\usepackage{fancyhdr}

\graphicspath{{./figures/}}

\def\BibTeX{{\rm B\kern-.05em{\sc i\kern-.025em b}\kern-.08em
    T\kern-.1667em\lower.7ex\hbox{E}\kern-.125emX}}

\renewcommand\st[1]{\unskip}

\newcommand{\etal}{\textit{et al.}}

\newcommand{\review}[1]{\textcolor{black}{#1}}

\begin{document}
\title{Transformer-based interpretable multi-modal data fusion for skin lesion classification}

\author{Theodor Cheslerean-Boghiu, Melia Fleischmann, Theresa Willem, and~\IEEEmembership{Tobias Lasser,~IEEE Member}
\thanks{This work was funded by the German Federal Ministry of Health’s program for digital innovations for the improvement of patient-centered care in healthcare [grant agreement no. 2520DAT920].}
\thanks{T. Cheslerean-Boghiu and T. Lasser are with the Group for Computational Imaging and Inverse Problems, Department of Informatics, School of Computation, Information, and Technology and Munich Institute of Biomedical Engineering at the Technical University of Munich, Germany (email: theo.cheslerean@tum.de, lasser@in.tum.de).}
\thanks{Melia Fleischmann is with the Department of Dermatology and Allergy, University Hospital, LMU Munich, Munich, Germany.}
\thanks{Theresa Willem is with the Institute of History and Ethics in Medicine, School of Medicine, and with the Department of Science, Technology and Society, School of Social Sciences and Technology, Technical
University of Munich, Munich, Germany.}
\thanks{For questions and inquiries please contact T. Cheslerean-Boghiu.}
}

\maketitle

\begin{abstract}
A lot of deep learning (DL) research these days is mainly focused on improving quantitative metrics regardless of other factors. In human-centered applications, like skin lesion classification in dermatology, DL-driven clinical decision support systems are still in their infancy due to the limited transparency of their decision-making process. Moreover, the lack of procedures that can explain the behavior of trained DL algorithms leads to almost no trust from clinical physicians. To diagnose skin lesions, dermatologists rely on visual assessment of the disease and the data gathered from the patient's anamnesis. Data-driven algorithms dealing with multi-modal data are limited by the separation of feature-level and decision-level fusion procedures required by convolutional architectures. To address this issue, we enable single-stage multi-modal data fusion via the attention mechanism of transformer-based architectures to aid in diagnosing skin diseases. Our method beats other state-of-the-art single- and multi-modal DL architectures in image-rich and patient-data-rich environments. Additionally, the choice of the architecture enables native interpretability support for the classification task both in the image and metadata domain with no additional modifications necessary.
\end{abstract}

\begin{IEEEkeywords}
data fusion, dermatology, interpretability, skin lesion classification, transformers, attention
\end{IEEEkeywords}

\thispagestyle{fancy}
\renewcommand{\headrulewidth}{0pt}
\fancyfoot[C]{\scriptsize{\copyright \copyright 2023 IEEE. Personal use of this material is permitted. Permission from IEEE must be obtained for all other uses, in any current or future media, including reprinting/republishing this material for advertising or promotional purposes, creating new collective works, for resale or redistribution to servers or lists, or reuse of any copyrighted component of this work in other works.}}

\section{Introduction} \label{section:introduction}

As the largest organ in the body, the skin serves as the first line of defense against a range of risks, including germs, viruses, moisture, and heat, among others. Additionally, it plays a critical role in immune system functions and helps regulate body temperature. Consequently, any emerging skin diseases should be promptly investigated by a specialized medical professional in skin, hair, and nail pathologies, such as a dermatologist. 

Critical to the diagnosis and treatment of over $\sim 3000$ skin lesions is the anamnesis of the patient. It is conducted during the initial consultation, when a dermatologist performs a visual inspection of the patient's skin and updates their medical history, recording any related symptoms in their report. Different pathology patterns are then considered based on the gathered information. To aid in the diagnosis process, clinicians often use established diagnostic procedures such as the ABCDE-rule~\cite{abcde-rule} or the Seven-Point Checklist~\cite{seven-point-criteria}. Additionally, they may conduct supplementary procedures such as blood tests, allergy testing, and biopsies to support the diagnosis \cite{braun-falco}.

These procedures allow clinicians to detect and evaluate cancerous cells, facilitating the timely provision of appropriate treatment options to patients. However, variations in the skin tumors' visual presentation (e.g. size, color, other physical characteristics) make the visual inspection time-consuming, prone to diagnostic errors, and reliant on the dermatologist's experience. Therefore, it is crucial to explore more precise and reliable diagnostic techniques that do not rely solely on visual assessment to reduce the risk of death in skin malignancies \cite{lookingbill-marks}.

\subsection{Background}
\subsubsection{Clinical Decision Support}
The success of deep learning in human-centered domains has prompted the rapid development of Computer-Aided Clinical Decision Support (CDS) systems as highly promising diagnostic aids in dermatology \cite{cds-assessment}. Deep learning (DL) algorithms are now applied in clinical practice through risk assessment models, enhancing diagnostic accuracy and workflow efficiency \cite{stellenwert}.

Several studies have shown that image-based DL-based algorithms are on par with some of their human counterparts in skin lesion image classification \cite{145dermatologists}. Even in the fields of psoriasis, neurodermatitis, or melanoma diagnostics, where the amount of labelled data is limited, classifiers have achieved notable performance \cite{limited-labelled-data}. However, for single-modality DL-based classifiers \cite{classres, classeff}, intra-class variance (shape, color, etc.), inter-class similarities (benign lesions that are similar to malign ones or vice versa), and the lack of available training data negatively influence the desired accuracy of diagnosis \cite{intrainter,deepmetric}.

In a clinical setting, dermatologists rely on information coming from multiple sources in different formats (dermoscopic image of the lesion, patient anamnesis and feedback, or a histology report from a pathologist) to make an informed diagnosis decision. Therefore, key to accurate DL-based skin lesion diagnosis is the design of DL algorithms that use fusion of the aforementioned multi-modal data sources to perform classification. The challenge of combining multi-modal data lies, however, in the scarcity of model architectures that can handle medically verifiable data fusion with sufficient performance compared to their human counter-parts \cite{dlmultimodalfusion}.

Addressing the data fusion challenge in our study, we chose various models to explore the performance of CDS when combining skin lesion image analysis with metadata and whether the outcomes could be made accessible and interpretable to medical experts.

\subsubsection{Multi-modal Data Fusion}
Early algorithms proposed by Ha~\etal~\cite{fusionconcat} and Yap~\etal~\cite{fusiondualconcat} combine the latent output of an image-fed Deep Convolutional Neural Network (DNN) via concatenation with the metadata either in its raw format or through its generated latent feature vector. A classification head then generates the classification output based on the concatenated latent features. Kawahara~\etal~\cite{multitaskmodal} used a triplet of one dermoscopic image, one clinical image obtained with a smartphone and a set of patient metadata to perform multi-label classification via feature-level fusion. Tang~\etal~\cite{fusionm4net} have noted that architectures enabling concatenation-based feature-level fusion of multi-modal data are suboptimal if the extracted latent feature vectors have mismatched sizes. Training a model under this setting leads to a biased model towards the feature-rich domain ignoring the rest.

Two-stage methods separate the feature-level from the decision-level fusion. Tang~\etal~\cite{fusionm4net} first performed image-based multi-label classification using the labels from the Seven-Point Checklist Dataset \cite{7pcdataset}. Then, the generated classification vectors are fused with the raw metadata vectors and SVMs are employed to perform decision-level fusion and subsequent classification. This approach has shown very promising results, however, by not performing any cross-modal feature-level fusion the classifier cannot develop any interaction between the different domains (image and metadata) which does not follow a standard dermatological diagnostic scenario.

While showing good performance, shallow DNNs fail to learn long-range contextual information, mainly due to the small receptive field of the convolution operation \cite{fullytransformer, vit, attall}. Chen~\etal~\cite{dilatcnn} attempted to overcome this limitation by increasing the receptive field of the convolutional layer using dilated convolutions, while Schlemper~\etal~\cite{attgate} developed a method similar to the attention functionality, where global dependencies are captured and employed against adversarial attacks with considerable success. Bi~\etal~\cite{hypercnn} proposed to perform the fusion in each convolutional layer throughout the network, thus connecting low-, mid-, and high-level image features with the metadata. We overcome the locality constraints of DNNs by using a transformer-based architecture to extract image-features and an attention layer to fuse the image and metadata latent vectors before the classification head (late fusion).

Attention-based architectures (transformers) are inherently able to learn mutual contextual information between all image regions and are not limited by any intrinsic local invariance as no convolution operations are used. Transformer models are capable of learning both semantic and positional information during training \cite{vit}, and they are data-agnostic at the cost of computational complexity. Additionally, the attention operation enables native interpretability, which makes transformers great candidates for medical multi-modal data fusion applications \cite{ontherelationship}.

Pacheco~\etal~\cite{metablock} have used a variant of the attention module to perform late data fusion in the latent space. In their network, metadata is used to produce a set of modifier coefficients to independently weigh the image features before the classifier. Cai~\etal~\cite{cainet} and Ou~\etal~\cite{dlmultimodalfusion} use the mechanism of co-attention (they call it \enquote{mutual attention}) to split the fusion step up into two branches, where on one branch the metadata attend to the image features, while the other branch performs the opposite task. Finally, they concatenate the output of the two attention layers and perform the classifications step. \review{The difference between using cross-attention instead of self-attention in the fusion step is that connections are only established between different modalities, regardless of any relationship within the querying modality. We propose to use self-attention on the concatenated multi-modal data, thus enabling the model to weigh the importance of each input token within the context of the entire input sequence.}

\review{More recently, multi-resolution transformer-based architectures have shown promising results over cross-modal or self-attention ones. Xu~\etal~\cite{remixformer} and Zhang~\etal~\cite{tformer} have proposed the RemixFormer and, respectively TFormer, which are both based on the Swin Transformer \cite{swintransformer} architecture. These models are composed of several SwinTransformer sub-layers that work in parallel at multiple patch size levels on the input. They have been shown to perform quite well on multi-modal image data datasets (dermatological and clinical images of the same skin lesion).}

All the abovementioned methods have been proposed to solve the skin lesion classification task via multi-modal data fusion. However, few of the current state-of-the-art works \review{\st{discuss the}provide an in-depth analysis} interpretability of their output as a CDS. We build on the current state-of-the-art by considering the implications as a decision support system of our proposed method using interpretability algorithms. \review{Additionally, some of the state-of-the-art architectures were designed for a particular use-case in dermatology, namely when, for each patient, multi-modal image data is also present (dermoscopic and non-dermoscopic images) besides categorical and text information. Many dermatological datasets, however, lack such diverse image data sources (ISIC 2018/2019 \cite{isic2019}, PAD UFES 20 \cite{padufes}, Fitspatrick-17k \cite{fitspatrick-17k-dataset}, SD-198 \cite{sd198}), nullifying the advantage of the aforementioned models if trained on them.}

\subsubsection{Interpretable AI}
Building trust between dermatologists and data-driven systems involves designing transparent models that explain \enquote{why they predict what they predict} \cite{gradcam}. Being able to interpret the decision-making process in the context of either a classification, segmentation, or denoising task is highly desirable in domains with impact on human lives. Selvaraju~\etal~\cite{gradcam} have discussed dividing the utility of inference transparency in three base categories: \textbf{failure identification} explains low performance of models, \textbf{confidence build-up} when the model matches the reasoning of humans (at the very least, or even improves on it), and \textbf{machine teaching} when the model outperforms human agents in the given metrics, and can explain the improvement.

In transformer-based architectures, the extracted attention maps from each layer represent a good indicator of support activations for the final classification, but they do not reflect the combined attention scores and the other components of the transformer model \cite{te}. Chefer~\etal~\cite{tmme} proposed a solution that integrates the gradient backpropagation with attention maps into one single model-wide relevancy map for multi-modal interactions with much better interpretability potential. Moreover, their method allows for the differentiation of positive and negative contributions to the decision output. 

We show in this work that the method proposed by Chefer~\etal~\cite{tmme} can be used to extract multi-modal interpretability information and that making such information accessible to dermatologists can increase the trust between the two parties.

\subsection{Contributions}
In this manuscript we propose a novel architecture that handles data fusion for skin lesion classification. Our contributions can be summarized as follows:
\begin{itemize}
	\item For multi-modal classification of skin lesions we \review{\st{combine}we adapt an existing architecture to combine} image features extracted with a transformer encoder \cite{vit} with embedded metadata features via a multi-head self-attention layer, and we call it VitAtt (\textit{vitatt}). The fused representation is then used for multi-class classification.
	\item We show that incorporating the multi-modal fusion process at feature-level has clear benefits over two-stage algorithms like \cite{fusionm4net} or competing transformers-based data fusion architectures \cite{cainet}.
	\item We show that the model can learn contextual information and learn correlation between domains. We propose to generate saliency maps for both image and metadata via the recently proposed Transformer Multi-Modal Explainability (TMME) method \cite{tmme}.
	\item We demonstrate that correct metadata feature engineering and training on a chosen subset of the metadata information can provide better classification performance.
	\item We show that the model learned certain aspects of the decision process that leads to a diagnosis by analyzing the metadata relevancy scores with medical doctors, and aligning the model's interpretation of the prediction to well-established diagnosis procedures.
	\item Lastly, we conduct a short analysis into the ethical considerations generated by the proposed architecture in combination with the used dataset, for which we provide a \textit{Model Card}~\cite{modelcard}.
\end{itemize}

\review{In Sec. \ref{section:methods} we introduce the data fusion architecture employed in this manuscript and describe the vision transformer model and the attention mechanism that it relies on. In Sec. \ref{section:experiments} we set up the environment for the experiments with the proposed method and the state-of-the-art, while in Sec. \ref{section:discussion} we review the results and discuss some of their implications from both technical and medical perspectives. Finally, in Sec. \ref{section:conclusion} we provide a short overview of the performed experiments and the results. Additionally, we compiled for the proposed architecture a Model Card \cite{modelcard} that contains additional information about the \textit{vitatt} model. The Supplementary Material contains new  experiments proving the performance of \textit{vitatt} over state-of-the-art models.}

\begin{figure}[t]
	\centering
	\includegraphics[width=0.75\linewidth]{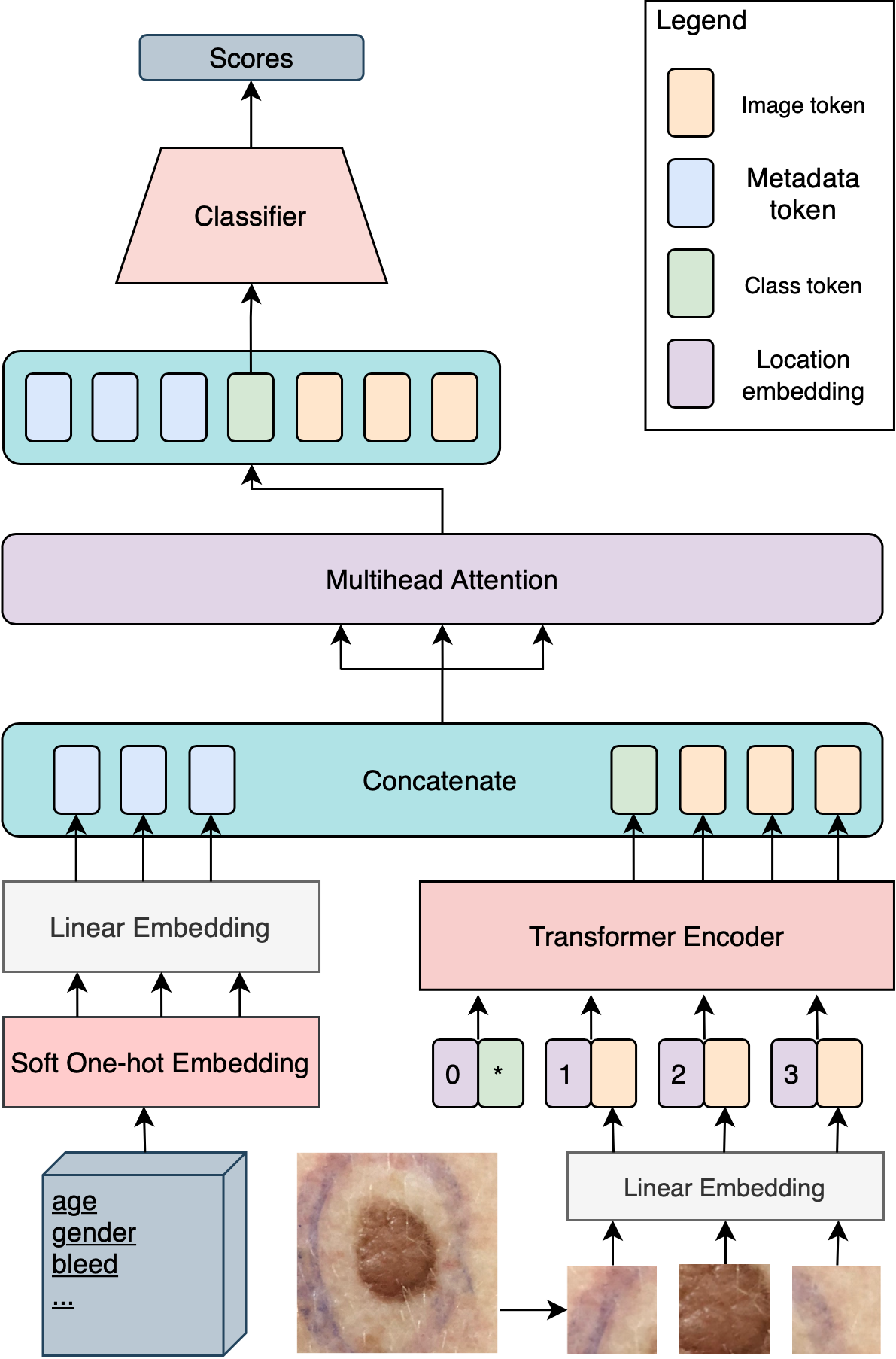}
	\caption{\small{Architecture of the proposed network for data fusion.}}
	\label{fig:architecture}
\end{figure}
\section{Methods} \label{section:methods}

Multi-class classification describes the task of classifying input instances into one of multiple classes. In multi-modal skin lesion classification, we seek to assign class scores to a pair composed of one lesion image and a set of patient metadata. The actual classification is then performed in two stages. The first stage relates to the feature extraction step in which a latent representation of the input is generated. Then, in the second step, the corresponding latent vectors are processed by a classification head into a set of non-normalized probabilities for each expected class. The class with the highest probability is then assumed to be the classification result.

Fig.~\ref{fig:architecture} shows the main structure of our proposed architecture. An input image is first separated into a sequence of patches (without overlap). Trainable positional embeddings added to each token enable the network to learn and retain positional information during training. One additional learnable token (\textbf{class}) is prepended to the input image patch set, whose state at the output of the transformer encoder serves as the image representation used by the classifier \cite{vit}. We employ a 12-layer transformer encoder (\textit{vit\_small\_patch16\_224}) to generate latent embeddings for each of the image patches and the \textbf{class} token.

Each layer of the transformer encoder contains a multi-head self-attention module. Compared to conventional convolutional layers, which work on the intrinsic assumption of locality and local invariance, attention layers allow for long-range dependencies to take shape by the application of a scaled dot product. Additionally, a fully connected layer, GeLU activation \cite{gelu} and layer normalization \cite{layernorm} are also used \footnote{\cite{timm} has additional details on the transformer encoder implementation.}.

The attention operation is defined as a function of three input arguments. A \enquote{query} $Q \in \mathbb{R}^{n}$ is multiplied via dot product with a \enquote{key} $K \in \mathbb{R}^{n}$ and the result is normalized by the square root of the key's dimension $\sqrt{d_k}$. The resulting weighting factor is then used to scale a \enquote{value} matrix $V \in \mathbb{R}^{n}$ and to generate the attention output. The attention pipeline is given by:
\begin{equation}
\text{Attention}(Q,K,V) = \text{softmax}\big( \frac{QK^T}{\sqrt{d_k}} \big) V
\end{equation}

In other words, the attention function is defined as the mapping of a query to a set of key-value pairs. It is beneficial that the query, key, and value are already located in the embedding space of the input, as the attention is then applied on the extracted features of the input \cite{attall}. 

To extract latent features from raw metadata into the shared latent space, we first encode the raw metadata values via soft one-hot encoding and then use one fully-connected layer coupled with a layer normalization operation to extract the metadata feature information. Afterwards, we concatenate the set of latent image tokens (including the \textbf{class} token) with the obtained latent metadata tokens in the channel dimension.

To enable data fusion based classification, we apply one final step of multi-head self-attention on the extended token set. This step enables the interaction between all image patches and metadata tokens under the influence of the \textbf{class} token, in contrast to co-attention modules where the query differs from the key and value inputs of the attention operation \cite{lxmert}. A residual connection is added to prevent degradation of the information fed through the fusion layer. 

Finally, the embedded \textbf{class} token after the fusion layer serves as the image representation used by a standard classification head composed of two fully connected layers, a batch normalization layer and \textit{Swish} activation \cite{swish}.

\section{Experiments} \label{section:experiments}

\begin{figure}[]
	\centering
	\begin{minipage}{0.40\linewidth}
		\centering
		\small
		Pre-Fusion
		\includegraphics[width=\linewidth]{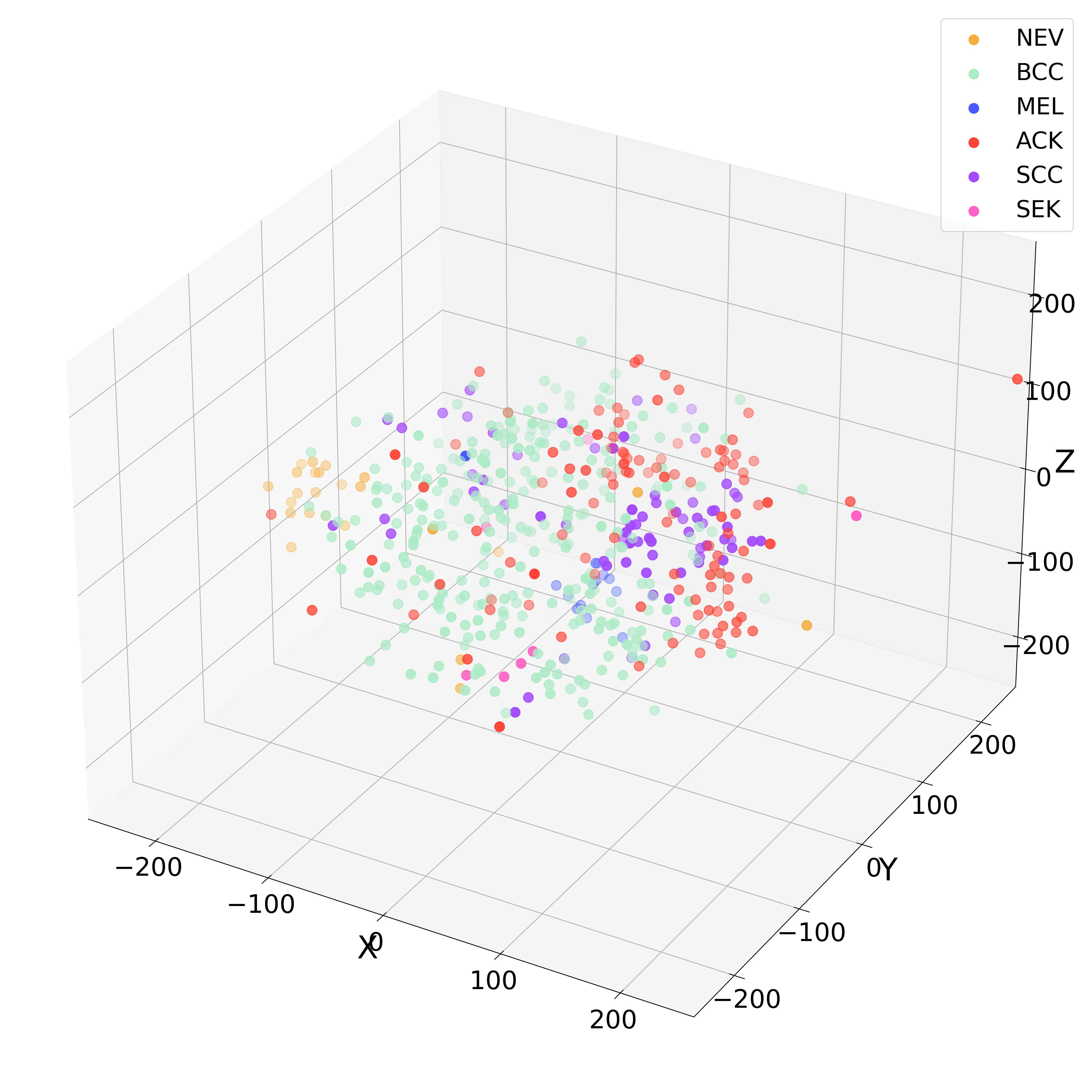}
	\end{minipage}
	\begin{minipage}{0.40\linewidth}
		\centering
		\small
		Post-Fusion
		\includegraphics[width=\linewidth]{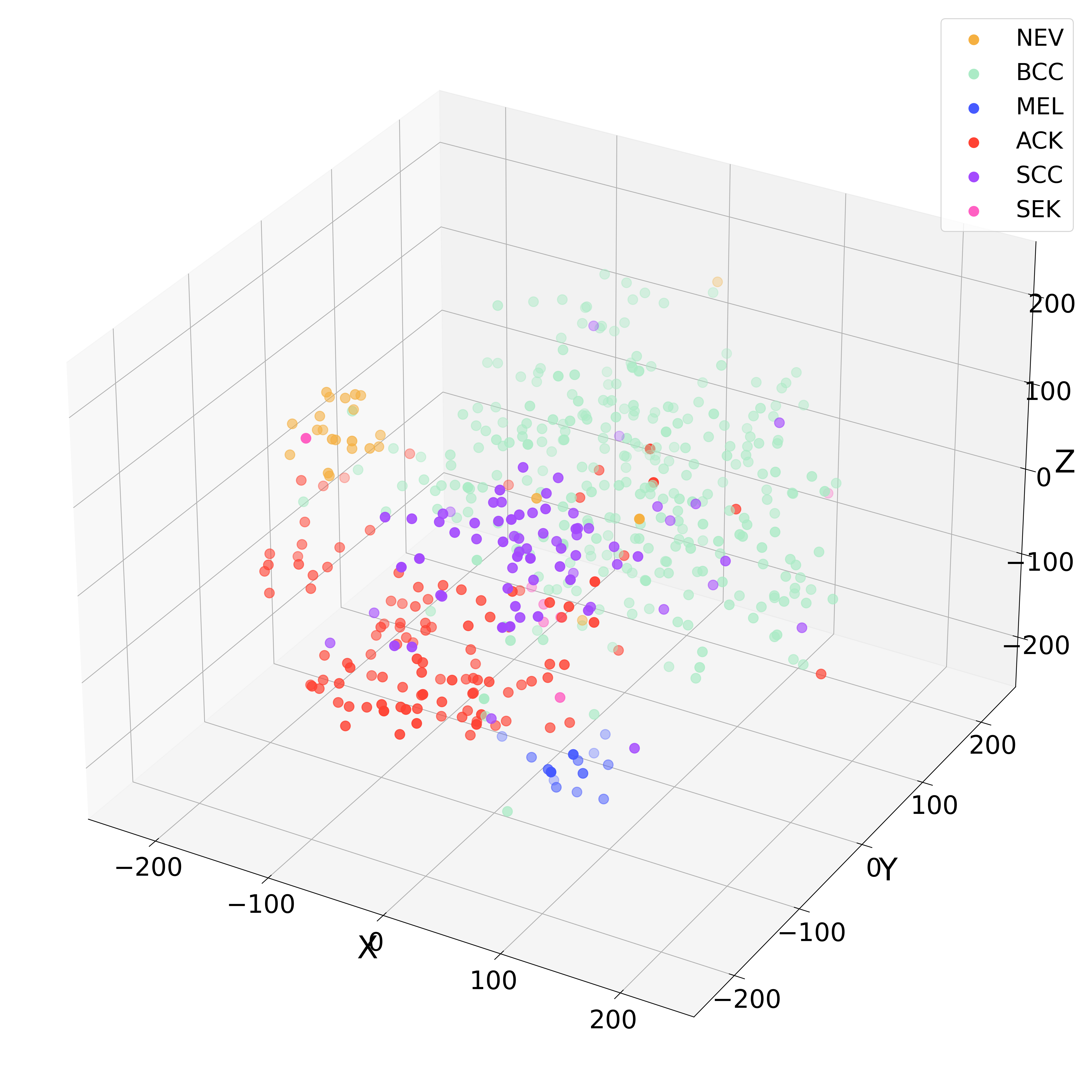}
	\end{minipage}
	\caption{\small{t-SNE based 3d visualization of the latent space prior and after the data fusion self-attention layer. Each scatter point denotes a sample from the test set. Clusters of different diseases are shown in different colors for Nevus (NEV), Basal Cell Carcinoma (BCC), Melanoma (MEL), Actinic Keratosis (ACK), Squamous Cell Carcinoma (SCC), Seborrheic Keratosis (SEK).}}
	\label{fig:latent-space}
\end{figure}

\subsection{Datasets}

\begin{figure}[h]
	\centering
	\includegraphics[width=0.85\linewidth]{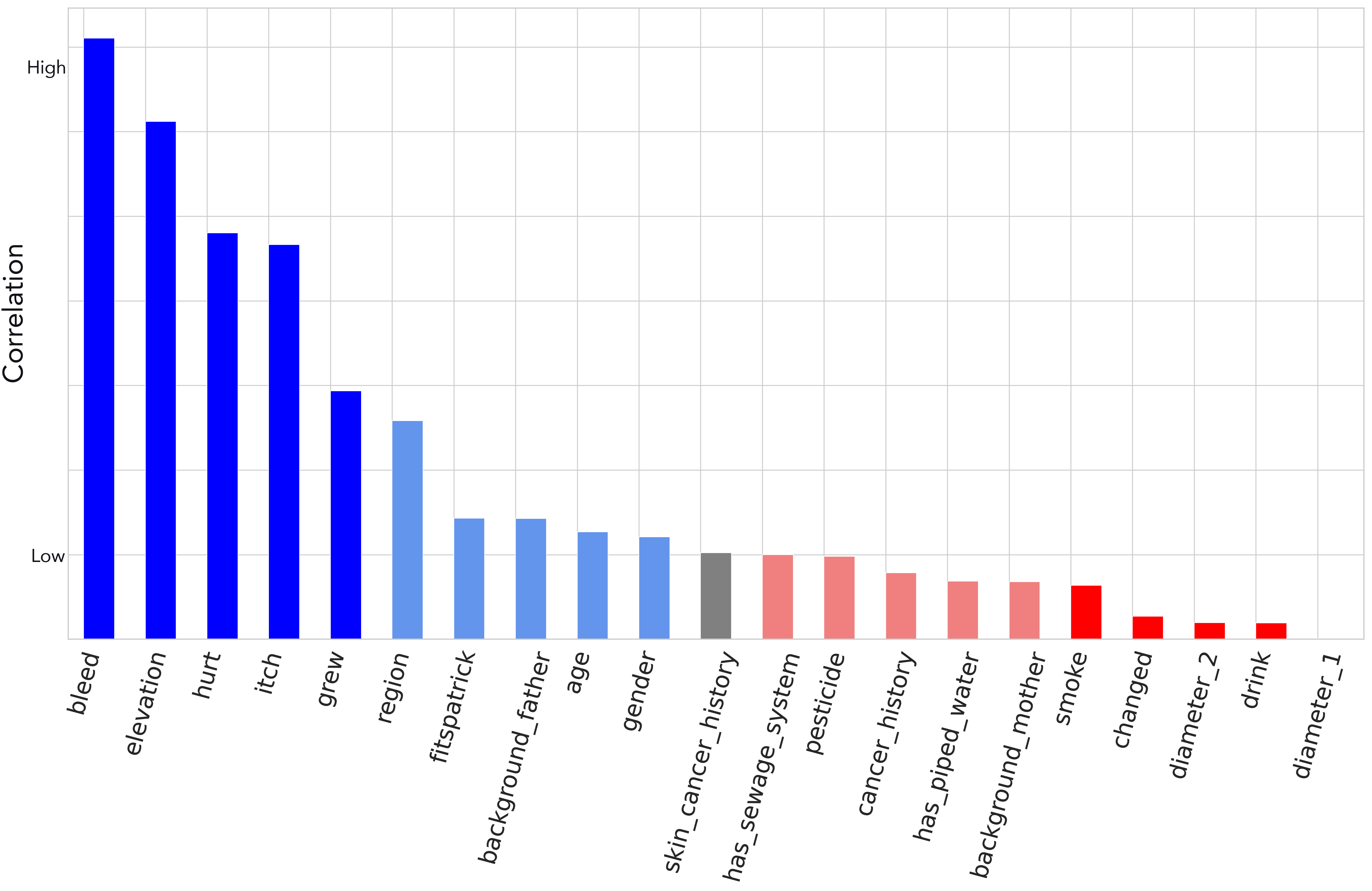}
	\caption{\small{Correlation coefficients ($y$-axis) computed between the \textit{diagnostic} label and the other metadata labels sorted in descending order. The HC-5, HC-10, LC-5, and LC-10 metadata labels used in Fig.~\ref{fig:bestworst-results} correspond to the blue(dark), blue(all), red(dark), red(all) bars.}}
	\label{fig:corrcoeff}
\end{figure}

PAD UFES 20 \cite{padufes} is a collection of clinical images of skin lesions (taken with a smartphone), coupled with individual patient metadata covering several topics of the patient anamnesis. In total there are $21$ metadata points for each person. Some of them describe whether the patient is a smoker (\textit{smoke}), has previously been diagnosed with cancer (\textit{cancer\_history}), skin type (as a \textit{fitspatrick} value), the birthplace of the parents (\textit{background\_mother}) or if the lesion showed signs of bleeding (\textit{bleed})\footnote{For a complete description of all metadata points check \cite{padufes}.}. One notable weakness of PAD UFES 20 is the mediocre quality of the contained clinical images compared to the high-quality dermoscopic ISIC 2019 images (see next paragraph), while one of its major strengths is the provided quantity of metadata for each patient. We first removed all image-metadata pairs from the dataset which contain undefined metadata points, and then partitioned the curated set ($\sim 1500$ samples remaining) into training, validation, and test subsets ($50\%/15\%/35\%$ ratio to provide sufficient samples for each class in the test set).

The ISIC 2019 dataset \cite{isic2019} has been developed as part of a challenge developed by the International Skin Imaging Collaboration (ISIC) with the goal to classify dermoscopic images into eight different diagnostic categories. The dataset consists of $\sim 25000$ images and corresponding patient metadata (\textit{age}, \textit{gender}, and \textit{anatomical location} of the lesion) available for training and testing across the eight classes, which makes it significantly bigger than PAD UFES 20.  We partitioned the ISIC-2019 training set into training, validation, and test subsets ($60\%/20\%/20\%$ ratio, same as in \cite{isiconline}).

The proportion of image- and metadata-information in these two datasets make them the perfect data source candidates for training and testing the different DL modalities, including our proposed one, as the performance of the models can be assessed both in a high-quality image-rich few metadata and a lower-quality image-poor rich metadata representation application domains.

\subsection{Evaluation metrics}
For performance evaluation we adopt five widely used performance metrics recommended in the ISIC Challenge: accuracy (ACC), sensitivity (SEN), specificity (SPE), precision (PRE), and area under the receiver operating characteristic curve (AUC) \cite{sensspec}. The \enquote{macro} average (avg) values of the metrics were used as main metrics for the performance comparisons.

\subsection{Comparison to state-of-the-art} \label{section:comparison}
\renewcommand{\arraystretch}{1}
\newcommand{\x}{0.05}
\definecolor{lightgray}{gray}{0.85}
\begin{figure*}[!h]
\centering
\begin{minipage}{0.49\linewidth}
    \footnotesize
	\begin{tabular}{| p{0.01\linewidth} | r | p{\x\linewidth} | p{\x\linewidth} | p{\x\linewidth} | p{\x\linewidth} | p{\x\linewidth} | p{\x\linewidth} | p{\x\linewidth} |}
	    \hline
	    & model & \bfseries NEV & \bfseries BCC & \bfseries MEL & \bfseries ACK & \bfseries SCC & \bfseries SEK & \bfseries avg \\\hline
	    & \# & 24 & 288 & 14 & 112 & 71 & 7 & \\\hline
	    \multirow{6}{*}{\rotatebox[origin=c]{90}{\footnotesize ACC}}&\textit{resnet} & 0.973 & 0.754 & 0.981 & 0.833 & 0.872 & 0.983 & 0.899 \\
	    &\textit{effnet} & 0.967 & 0.752 & \bfseries 0.990 & 0.816 & 0.866 & \bfseries 0.988 & 0.897 \\
	    &\textit{vit} & 0.983 & 0.766 & 0.984 & 0.860 & 0.876 & \bfseries 0.988 & 0.910 \\
	    &\textbf{\textit{vitatt}} & \bfseries 0.986 & \bfseries 0.841 & \bfseries 0.990 & \bfseries 0.915 & \bfseries 0.890 & 0.986 & \bfseries 0.935 \\
	    &\textit{cainet} & 0.981 & 0.785 & 0.988 & 0.876 & 0.884 & 0.986 & 0.917 \\
	    &\textit{fm4net} & 0.981 & 0.758 & 0.983 & 0.851 & 0.880 & 0.986 & 0.906 \\

        \bottomrule

        \multirow{6}{*}{\rotatebox[origin=c]{90}{\footnotesize PRE}}&\textit{resnet} & 0.667 & 0.762 & 0.667 & 0.612 & 0.558 & 0.375 & 0.607\\
	    &\textit{effnet} & 0.606 & 0.755 & \bfseries 0.909 & 0.581 & 0.521 & \bfseries 0.600 & 0.662\\
	    &\textit{vit} & 0.826 & 0.738 & 0.688 & 0.717 & 0.621 & \bfseries 0.600 & 0.698\\
	    &\textbf{\textit{vitatt}} & \bfseries 0.870 & \bfseries 0.808 & 0.800 & \bfseries 0.854 & 0.667 & 0.500 & \bfseries 0.750\\
	    &\textit{cainet} & 0.818 & 0.752 & 0.833 & 0.750 & \bfseries 0.704 & 0.500 & 0.726\\
	    &\textit{fm4net} & 0.792 & 0.738 & 0.727 & 0.677 & 0.636 & 0.500 & 0.678\\

        \bottomrule

        \multirow{6}{*}{\rotatebox[origin=c]{90}{\footnotesize SEN}}&\textit{resnet} & \bfseries 0.833 & 0.812 & 0.571 & 0.634 & 0.338 & 0.429 & 0.603\\
        &\textit{effnet} & \bfseries 0.833 & 0.823 & 0.714 & 0.545 & 0.352 & 0.429 & 0.616\\
        &\textit{vit} & 0.792 & 0.899 & 0.786 & 0.589 & 0.254 & 0.429 & 0.625\\
        &\textbf{\textit{vitatt}} & \bfseries 0.833 & \bfseries 0.938 & \bfseries 0.857 & \bfseries 0.732 & \bfseries 0.394 & 0.429 & \bfseries 0.697\\
        &\textit{cainet} & 0.750 & 0.917 & 0.714 & 0.643 & 0.268 & \bfseries 0.571 & 0.644\\
        &\textit{fm4net} & 0.792 & 0.878 & 0.571 & 0.598 & 0.296 & 0.429 & 0.594\\
        
        \bottomrule

        \multirow{6}{*}{\rotatebox[origin=c]{90}{\footnotesize SPE}}&\textit{resnet} & 0.980 & 0.680 & 0.992 & 0.889 & 0.957 & 0.990 & 0.915\\
        &\textit{effnet} & 0.974 & 0.662 & \bfseries 0.998 & 0.891 & 0.948 & \bfseries 0.996 & 0.912\\
        &\textit{vit} & 0.992 & 0.596 & 0.990 & 0.936 & 0.975 & \bfseries 0.996 & 0.914\\
        &\textbf{\textit{vitatt}} & \bfseries 0.994 & \bfseries 0.719 & 0.994 & \bfseries 0.965 & 0.969 & 0.994 & \bfseries 0.939\\
        &\textit{cainet} & 0.992 & 0.618 & 0.996 & 0.941 & \bfseries 0.982 & 0.992 & 0.920\\
        &\textit{fm4net} & 0.990 & 0.605 & 0.994 & 0.921 & 0.973 & 0.994 & 0.913\\

        \bottomrule

        \multirow{6}{*}{\rotatebox[origin=c]{90}{\footnotesize SPE}}&\textit{resnet} & 0.907 & 0.746 & 0.782 & 0.761 & 0.648 & 0.709 & 0.759\\
        &\textit{effnet} & 0.903 & 0.743 & 0.856 & 0.718 & 0.650 & 0.712 & 0.764\\
        &\textit{vit} & 0.892 & 0.748 & 0.888 & 0.762 & 0.614 & 0.712 & 0.769\\
        &\textbf{\textit{vitatt}} & \bfseries 0.914 & \bfseries 0.828 & \bfseries 0.926 & \bfseries 0.849 & \bfseries 0.681 & 0.711 & \bfseries 0.818\\
        &\textit{cainet} & 0.871 & 0.768 & 0.855 & 0.792 & 0.625 & \bfseries 0.782 & 0.782\\
        &\textit{fm4net} & 0.891 & 0.742 & 0.783 & 0.760 & 0.634 & 0.711 & 0.753\\\hline
    \end{tabular}
\end{minipage}
\begin{minipage}{0.49\linewidth}
\centering
    \footnotesize
	\begin{tabular}{| p{0.01\linewidth} | r | p{\x\linewidth} | p{\x\linewidth} | p{\x\linewidth} | p{\x\linewidth} | p{\x\linewidth} | p{\x\linewidth} | p{\x\linewidth} | p{\x\linewidth} | p{\x\linewidth} |}
	    \toprule
	    & model  & \bfseries NEV & \bfseries BCC & \bfseries MEL & \bfseries ACK & \bfseries SCC & \bfseries BKL & \bfseries VASC & \bfseries DF & \bfseries avg \\\hline
	    & \# & 2002 & 602 & 821 & 157 & 121 & 469 & 46 & 44 & \\

    \bottomrule

    \multirow{6}{*}{\rotatebox[origin=c]{90}{\footnotesize ACC}}&\textit{resnet} & 0.920 & 0.970 & 0.924 & 0.982 & 0.988 & 0.956 & 0.998 & 0.997 & 0.967\\
	&\textit{effnet} & 0.941 & 0.973 & 0.947 & \bfseries 0.985 & \bfseries 0.989 & 0.962 & 0.998 & 0.996 & 0.974\\
	&\textit{vit}    & 0.930 & 0.965 & 0.932 & 0.980 & \bfseries 0.989 & 0.948 & 0.999 & 0.997 & 0.967\\
	& \textbf{\textit{vitatt}} & \bfseries 0.948 & \bfseries 0.975 & \bfseries 0.954 & \bfseries 0.985 & \bfseries 0.989 & \bfseries 0.964 & \bfseries 0.999 & \bfseries 0.998 & \bfseries 0.976\\
	&\textit{cainet} & 0.925 & 0.969 & 0.932 & \bfseries 0.985 & 0.987 & 0.955 & 0.998 & \bfseries 0.998 & 0.969\\
	&\textit{fm4net} & 0.920 & 0.970 & 0.924 & 0.982 & 0.988 & 0.956 & 0.998 & 0.997 & 0.967\\

    \bottomrule

    \multirow{6}{*}{\rotatebox[origin=c]{90}{\footnotesize PRE}}&\textit{resnet} & 0.883 & 0.894 & 0.895 & 0.738 & 0.818 & 0.788 & 0.896 & 0.841 & 0.844\\
	&\textit{effnet} & 0.913 & 0.893 & \bfseries 0.913 & 0.788 & \bfseries 0.857 & \bfseries 0.845 & \bfseries 0.976 & 0.795 & 0.873\\
	&\textit{vit}    & 0.906 & 0.858 & 0.874 & 0.708 & 0.819 & 0.779 & 0.956 & 0.851 & 0.844\\
	&\textbf{\textit{vitatt}} & \bfseries 0.929 & \bfseries 0.902 & 0.908 & 0.806 & 0.852 & \bfseries 0.845 & 0.917 & 0.929 & \bfseries 0.886\\
	&\textit{cainet} & 0.898 & 0.862 & 0.898 & \bfseries 0.831 & 0.754 & 0.792 & 0.915 & \bfseries 1.000 & 0.869\\
	&\textit{fm4net} & 0.883 & 0.894 & 0.895 & 0.738 & 0.818 & 0.788 & 0.896 & 0.841 & 0.844\\

    \bottomrule

    \multirow{6}{*}{\rotatebox[origin=c]{90}{\footnotesize SEN}}&\textit{resnet} & 0.956 & 0.895 & 0.686 & 0.809 & 0.744 & \bfseries 0.823 & 0.935 & 0.841 & 0.836\\
	&\textit{effnet} & \bfseries 0.967 & 0.919 & 0.804 & \bfseries 0.828 & 0.744 & 0.802 & 0.870 & 0.795 & 0.841\\
	&\textit{vit}    & 0.949 & 0.900 & 0.758 & 0.771 & 0.785 & 0.736 & 0.935 & \bfseries 0.909 & 0.843\\
	&\textbf{\textit{vitatt}} & 0.963 & 0.922 & \bfseries 0.844 & 0.796 & 0.760 & \bfseries 0.823 & \bfseries 0.957 & 0.886 & \bfseries 0.869\\
	&\textit{cainet} & 0.948 & \bfseries 0.934 & 0.731 & 0.752 & \bfseries 0.810 & 0.802 & 0.935 & 0.773 & 0.835\\
	&\textit{fm4net} & 0.956 & 0.895 & 0.686 & 0.809 & 0.744 & \bfseries 0.823 & 0.935 & 0.841 & 0.836\\

    \bottomrule

    \multirow{6}{*}{\rotatebox[origin=c]{90}{\footnotesize SPE}}&\textit{resnet} & 0.888 & 0.983 & 0.981 & 0.989 & 0.995 & 0.973 & 0.999 & 0.998 & 0.976\\
	&\textit{effnet} & 0.918 & 0.982 & \bfseries 0.982 & 0.991 & \bfseries 0.996 & 0.979 & \bfseries 1.000 & 0.998 & 0.981\\
	&\textit{vit}    & 0.913 & 0.975 & 0.974 & 0.988 & 0.995 & 0.974 & \bfseries 1.000 & 0.998 & 0.977\\
	&\textbf{\textit{vitatt}} & \bfseries 0.935 & \bfseries 0.984 & 0.980 & 0.993 & \bfseries 0.996 & \bfseries 0.981 & \bfseries 1.000 & 0.999 & \bfseries 0.983\\
	&\textit{cainet} & 0.904 & 0.975 & 0.980 & \bfseries 0.994 & 0.992 & 0.974 & 0.999 & \bfseries 1.000 & 0.977\\
	&\textit{fm4net} & 0.888 & 0.983 & 0.981 & 0.989 & 0.995 & 0.973 & 0.999 & 0.998 & 0.976\\

    \bottomrule

    \multirow{6}{*}{\rotatebox[origin=c]{90}{\footnotesize AUC}}&\textit{resnet} & 0.922 & 0.939 & 0.833 & 0.899 & 0.869 & 0.898 & 0.967 & 0.920 & 0.906\\
	&\textit{effnet} & 0.942 & 0.950 & 0.893 & \bfseries 0.910 & 0.870 & 0.892 & 0.935 & 0.897 & 0.911\\
	&\textit{vit}    & 0.931 & 0.938 & 0.866 & 0.879 & 0.890 & 0.855 & 0.967 & \bfseries 0.954 & 0.910\\
	&\textbf{\textit{vitatt}} & \bfseries 0.949 & 0.953 & \bfseries 0.912 & 0.894 & 0.878 & \bfseries 0.902 & \bfseries 0.978 & 0.943 & \bfseries 0.926\\
	&\textit{cainet} & 0.926 & \bfseries 0.954 & 0.856 & 0.873 & \bfseries 0.901 & 0.888 & 0.967 & 0.886 & 0.906\\
	&\textit{fm4net} & 0.922 & 0.939 & 0.833 & 0.899 & 0.869 & 0.898 & 0.967 & 0.920 & 0.906\\

    \bottomrule

	\end{tabular}
\end{minipage}

\caption{\small{Quantitative results on PAD UFES 20 (\textbf{left}) and ISIC 2019 (\textbf{right}) for both image-based and multi-modal multi-class classification. The first three rows represent the single-modality trained networks: \textit{resnet} (ResNet-50), \textit{effnet} (EfficientNet-b0), \textit{vit} (Vision Transformer). The last three rows represent the multi-modality trained networks: \textit{vitatt} (ours, highlighted by the gray row), \textit{cainet} (Cai et al.~\cite{cainet}), \textit{fm4net} (FusionM4Net~\cite{fusionm4net}). The single-modality networks \textit{resnet} and \textit{vit} correspond to the image feature extractors from \textit{fm4net} and \textit{vitatt}. This allows for a direct comparison between image-based training and multi-modal training. The highest metric value for intra-class and average metrics has been marked in bold font.}}

\label{fig:main-results}
\end{figure*}

For performance comparison we use state-of-the art models, as introduced in Sec.~\ref{section:introduction}. We train two convolutional models and one vision transformer on image-data only: ResNet-50 (\textit{resnet}) \cite{resnet}, EfficientNet-b0 (\textit{effnet}) \cite{efficientnet}, and ViT (\textit{vit}) \cite{vit}. The ViT model is the \textit{vit\_small\_patch16\_224} model used by the \textit{timm} library \cite{timm}\footnote{The size of the query, key, value tensors is in this case $n=384$.}. All three models were pretrained on the ImageNet dataset. The other two networks are CaiNet (\textit{cainet}) \cite{cainet} and FusionM4Net (\textit{fm4net}) \cite{fusionm4net}, which are state-of-the-art data fusion architectures for skin lesion classification. These models were mainly chosen thanks to their different approach to data fusion. \textit{cainet} performs feature-level fusion in the latent space via the co-attention mechanism, while \textit{fm4net} fuses the multi-modal data in a two-stage architecture, where the first stage handles the image classification and the second stage fine-tunes the classification scores with the help of additional patient metadata (decision-level fusion). The image-domain branch of these networks was pretrained on the ImageNet dataset as was the image feature extractor in our network. \textit{vitatt} does not use \textit{vit} as a warm-start for the image feature extractor. \review{One additional network TFormer (\textit{tformer} \cite{tformer}), which has been recently published, has been also trained and evaluated together with the rest of the models. The Supplementary Material section of this manuscript contains the setup and results for the statistical experiments that we conducted to validate the superior performance of \textit{vitatt} over the three other comparable data fusion models (\textit{cainet}, \textit{fm4net}, \textit{tformer}).}

It has already been shown in the literature that the \textit{effnet}-based architectures outperform the \textit{resnet}- and \textit{vit}-based ones \cite{efficientnet}, however we added the latter two as they are used as backbones in the \textit{fm4net} and \textit{vitatt} (our) models. Their performance metrics are thus used to show the improvement of data fusion over conventional single-domain image-based classification.

Fig.~\ref{fig:main-results} shows the quantitative results of the comparisons on the PAD UFES 20 and the ISIC 2019 dataset. Additionally, the \enquote{macro} average of these metrics was added on each row. 

We also performed an investigation into the balance and number of metadata available to the network during training. Using the correlation coefficient computed on the training set between all metadata labels and the \textit{diagnostic} label (seen in Fig.~\ref{fig:corrcoeff}) we picked five and ten labels with the \textbf{h}ighest corresponding \textbf{c}orrelation coefficient (HC-5 and HC-10) and five and ten with the \textbf{l}owest \textbf{c}orrelation coefficients (LC-5 and LC-10). We then trained three networks: \textit{vitatt}, \textit{cainet} and \textit{fm4net} with the four metadata sets, while keeping the training set size fixed, and plotted in Fig.~\ref{fig:bestworst-results} the quantitative evaluation metrics on the test set. The evaluation of the trained network was performed using the same metadata sets.

\begin{figure}[]
\centering
	\begin{minipage}{\linewidth}
	\centering
	    \footnotesize
		\begin{tabular}{| p{0.01\linewidth} | p{0.01\linewidth} | r |p{\x\linewidth} |p{\x\linewidth} | p{\x\linewidth} | p{\x\linewidth} | p{\x\linewidth} | p{\x\linewidth} | p{\x\linewidth} |}
		    \toprule
		    & & model & \bfseries NEV & \bfseries BCC & \bfseries MEL & \bfseries ACK & \bfseries SCC & \bfseries SEK & \bfseries avg \\\hline
		    & & \# & 24 & 288 & 14 & 112 & 71 & 7 & \\\hline
            \multirow{12}{*}{\rotatebox[origin=c]{90}{\footnotesize ACC}} & \multirow{3}{*}{\rotatebox[origin=c]{90}{LC-5}}
            &\textbf{\textit{vitatt}} & \bfseries 0.979 & 0.773 & \bfseries 0.986 & \bfseries 0.880 & 0.859 & 0.981 & 0.910\\
            & &\textit{cainet} & 0.975 & 0.767 & \bfseries 0.986 & 0.874 & \bfseries 0.878 & \bfseries 0.984 & \bfseries 0.911\\
            & &\textit{fm4net} & 0.941 & 0.719 & 0.977 & 0.828 & 0.833 & 0.977 & 0.879\\ \cmidrule{2-10}
            &\multirow{3}{*}{\rotatebox[origin=c]{90}{LC-10}}
            & \textbf{\textit{vitatt}} & \bfseries 0.986 & 0.766 & \bfseries 0.984 & 0.845 & 0.860 & \bfseries 0.988 & \bfseries 0.905\\
            & &\textit{cainet} & 0.963 & \bfseries 0.775 & 0.975 & \bfseries 0.851 & \bfseries 0.888 & 0.959 & 0.902\\
            & &\textit{fm4net} & 0.981 & 0.752 & 0.981 & 0.839 & 0.886 & 0.984 & 0.904\\ \cmidrule{2-9}
            &\multirow{3}{*}{\rotatebox[origin=c]{90}{HC-5}}
            & \textbf{\textit{vitatt}} & \bfseries 0.990 & 0.795 & \bfseries 0.988 & 0.893 & 0.870 & 0.983 & \bfseries 0.920\\
            & &\textit{cainet} & 0.973 & \bfseries 0.804 & 0.979 & \bfseries 0.903 & \bfseries 0.874 & \bfseries 0.986 & \bfseries 0.920\\
            & &\textit{fm4net} & 0.983 & 0.750 & 0.981 & 0.839 & 0.882 & 0.984 & 0.903\\ \cmidrule{2-9}
            &\multirow{3}{*}{\rotatebox[origin=c]{90}{HC-10}}
            &\textbf{\textit{vitatt}} & \bfseries 0.983 & 0.797 & \bfseries 0.988 & \bfseries 0.905 & 0.876 & \bfseries 0.990 & \bfseries 0.923\\
            & &\textit{cainet} & 0.973 & \bfseries 0.818 & 0.983 & 0.859 & \bfseries 0.878 & \bfseries 0.990 & 0.917\\
            & &\textit{fm4net} & 0.983 & 0.752 & 0.981 & 0.843 & 0.880 & 0.984 & 0.904\\ 

            \bottomrule

            \multirow{12}{*}{\rotatebox[origin=c]{90}{\footnotesize PRE}} & \multirow{3}{*}{\rotatebox[origin=c]{90}{LC-5}}
            &\textbf{\textit{vitatt}} & \bfseries 0.783 & \bfseries 0.741 & \bfseries 0.889 & 0.784 & 0.471 & 0.286 & 0.659\\
            & &\textit{cainet} & 0.690 & 0.736 & 0.889 & \bfseries 0.813 & \bfseries 0.591 & \bfseries 0.333 & \bfseries 0.675\\
            & &\textit{fm4net} & 0.357 & 0.730 & 0.667 & 0.486 & 0.429 & 0.800 & 0.578\\ \cmidrule{2-10}
            & \multirow{3}{*}{\rotatebox[origin=c]{90}{LC-10}}
            &\textbf{\textit{vitatt}} & \bfseries 0.870 & 0.742 & \bfseries 0.750 & \bfseries 0.663 & 0.485 & \bfseries 0.600 & \bfseries 0.685\\
            & &\textit{cainet} & 0.600 & \bfseries 0.777 & 0.667 & 0.650 & \bfseries 0.667 & 0.182 & 0.590\\
            & &\textit{fm4net} & 0.792 & 0.735 & 0.700 & 0.641 & 0.688 & 0.429 & 0.664\\ \cmidrule{2-10}
            & \multirow{3}{*}{\rotatebox[origin=c]{90}{HC-5}}
            &\textbf{\textit{vitatt}} & \bfseries 0.913 & 0.781 & \bfseries 1.000 & 0.721 & \bfseries 0.583 & 0.375 & \bfseries 0.729\\
            & &\textit{cainet} & 0.812 & \bfseries 0.788 & 0.600 & \bfseries 0.798 & 0.558 & \bfseries 0.500 & 0.676\\
            & &\textit{fm4net} & 0.826 & 0.733 & 0.700 & 0.641 & 0.656 & 0.429 & 0.664\\ \cmidrule{2-10}
            & \multirow{3}{*}{\rotatebox[origin=c]{90}{HC-10}}
            &\textbf{\textit{vitatt}} & \bfseries 0.826 & 0.765 & \bfseries 0.833 & \bfseries 0.839 & \bfseries 0.590 & \bfseries 0.750 & \bfseries 0.767\\
            & &\textit{cainet} & 0.727 & \bfseries 0.821 & 0.727 & 0.654 & 0.580 & 0.750 & 0.710\\
            & &\textit{fm4net} & 0.826 & 0.733 & 0.700 & 0.657 & 0.636 & 0.429 & 0.663\\

            \bottomrule

            \multirow{12}{*}{\rotatebox[origin=c]{90}{\footnotesize SEN}} & \multirow{3}{*}{\rotatebox[origin=c]{90}{LC-5}}
            &\textbf{\textit{vitatt}} & 0.750 &\bfseries  0.913 & \bfseries 0.571 & \bfseries 0.616 & 0.225 & \bfseries 0.286 & 0.560\\
            & &\textit{cainet} & \bfseries 0.833 & 0.910 & \bfseries 0.571 & 0.545 & \bfseries 0.366 & 0.143 & \bfseries 0.561\\
            & &\textit{fm4net} & 0.556 & 0.800 & 0.750 & 0.459 & 0.265 & 0.500 & 0.555\\ \cmidrule{2-10}
            & \multirow{3}{*}{\rotatebox[origin=c]{90}{LC-10}}
            &\textbf{\textit{vitatt}} & \bfseries 0.833 & \bfseries 0.889 & \bfseries 0.643 & 0.580 & 0.225 & 0.429 & \bfseries 0.600\\
            & &\textit{cainet} & 0.625 & 0.837 & 0.143 & \bfseries 0.679 & \bfseries 0.366 & \bfseries 0.571 & 0.537\\
            & &\textit{fm4net} & 0.792 & 0.868 & 0.500 & 0.589 & 0.310 & 0.429 & 0.581\\ \cmidrule{2-10}
            & \multirow{3}{*}{\rotatebox[origin=c]{90}{HC-5}}
            &\textbf{\textit{vitatt}} & \bfseries 0.875 & 0.878 & 0.571 & \bfseries 0.830 & 0.197 & \bfseries 0.429 & \bfseries 0.630\\
            & &\textit{cainet} & 0.542 & \bfseries 0.889 & \bfseries 0.643 & 0.741 & \bfseries 0.408 & 0.286 & 0.585\\
            & &\textit{fm4net} & 0.792 & 0.868 & 0.500 & 0.589 & 0.296 & 0.429 & 0.579\\ \cmidrule{2-10}
            & \multirow{3}{*}{\rotatebox[origin=c]{90}{HC-10}}
            &\textbf{\textit{vitatt}} & \bfseries 0.792 & \bfseries 0.917 & \bfseries 0.714 & 0.696 & 0.324 & \bfseries 0.429 & \bfseries 0.645\\
            & &\textit{cainet} & 0.667 & 0.861 & 0.571 & \bfseries 0.741 & \bfseries 0.408 & \bfseries 0.429 & 0.613\\
            & &\textit{fm4net} & 0.792 & 0.875 & 0.500 & 0.580 & 0.296 & 0.429 & 0.579\\

            \bottomrule

            \multirow{12}{*}{\rotatebox[origin=c]{90}{\footnotesize SPE}} & \multirow{3}{*}{\rotatebox[origin=c]{90}{LC-5}}
            &\textbf{\textit{vitatt}} & \bfseries 0.979 & \bfseries 0.773 & \bfseries 0.986 & \bfseries 0.880 & 0.859 & 0.981 & 0.910\\
            & &\textit{cainet} & 0.975 & 0.767 & \bfseries 0.986 & 0.874 & \bfseries 0.878 & \bfseries 0.984 & \bfseries 0.911\\
            & &\textit{fm4net} & 0.941 & 0.719 & 0.977 & 0.828 & 0.833 & 0.977 & 0.879\\ \cmidrule{2-10}
            & \multirow{3}{*}{\rotatebox[origin=c]{90}{LC-10}}
            &\textbf{\textit{vitatt}} & \bfseries 0.994 & 0.610 & 0.994 & \bfseries 0.918 & 0.962 & \bfseries 0.996 & 0.912\\
            & &\textit{cainet} & 0.980 & \bfseries 0.697 & \bfseries 0.998 & 0.899 & 0.971 & 0.965 & \bfseries 0.918\\
            & &\textit{fm4net} & 0.990 & 0.605 & 0.994 & 0.908 & \bfseries 0.978 & 0.992 & 0.911\\ \cmidrule{2-10}
            & \multirow{3}{*}{\rotatebox[origin=c]{90}{HC-5}}
            &\textbf{\textit{vitatt}} & \bfseries 0.996 & 0.689 & 1.000 & 0.911 & \bfseries 0.978 & 0.990 & 0.927\\
            & &\textit{cainet} & 0.994 & \bfseries 0.697 & \bfseries 0.988 & \bfseries 0.948 & 0.948 & \bfseries 0.996 & \bfseries \bfseries 0.929\\
            & &\textit{fm4net} & 0.992 & 0.601 & 0.994 & 0.908 & 0.975 & 0.992 & 0.910\\ \cmidrule{2-10}
            & \multirow{3}{*}{\rotatebox[origin=c]{90}{HC-10}}
            &\textbf{\textit{vitatt}} & \bfseries 0.992 & 0.645 & \bfseries 0.996 & \bfseries 0.963 & 0.964 & \bfseries 0.998 & 0.926\\
            & &\textit{cainet} & 0.988 & \bfseries 0.763 & 0.994 & 0.891 & 0.953 &\bfseries  0.998 & \bfseries 0.931\\
            & &\textit{fm4net} & \bfseries 0.992 & 0.596 & 0.994 & 0.916 & \bfseries 0.973 & 0.992 & 0.911\\

            \bottomrule

            \multirow{12}{*}{\rotatebox[origin=c]{90}{\footnotesize AUC}} & \multirow{3}{*}{\rotatebox[origin=c]{90}{LC-5}}
            &\textbf{\textit{vitatt}} & 0.870 & \bfseries 0.755 & \bfseries 0.785 & \bfseries 0.785 & 0.592 & 0.638 & 0.737\\
            & &\textit{cainet} & \bfseries 0.908 & 0.749 & \bfseries 0.785 & 0.755 & \bfseries 0.663 & 0.569 & \bfseries 0.738\\
            & &\textit{fm4net} & 0.757 & 0.707 & 0.868 & 0.681 & 0.600 & \bfseries 0.748 & 0.727\\ \cmidrule{2-10}
            & \multirow{3}{*}{\rotatebox[origin=c]{90}{LC-10}}
            &\textbf{\textit{vitatt}} & \bfseries 0.914 & 0.749 & \bfseries 0.818 & 0.749 & 0.594 & 0.712 & \bfseries 0.756\\
            & &\textit{cainet} & 0.802 & \bfseries 0.767 & 0.570 & \bfseries 0.789 & \bfseries 0.668 & \bfseries 0.768 & 0.727\\
            & &\textit{fm4net} & 0.891 & 0.737 & 0.747 & 0.749 & 0.644 & 0.710 & 0.746\\ \cmidrule{2-10}
            & \multirow{3}{*}{\rotatebox[origin=c]{90}{HC-5}}
            &\textbf{\textit{vitatt}} & \bfseries 0.935 & 0.784 & 0.786 & \bfseries 0.871 & 0.587 & 0.709 & \bfseries 0.779\\
            & &\textit{cainet} & 0.768 & \bfseries 0.793 & \bfseries 0.815 & 0.845 & \bfseries 0.678 & 0.641 & 0.757\\
            & &\textit{fm4net} & 0.892 & 0.734 & 0.747 & 0.749 & 0.636 & \bfseries 0.710 & 0.745\\ \cmidrule{2-10}
            & \multirow{3}{*}{\rotatebox[origin=c]{90}{HC-10}}
            &\textbf{\textit{vitatt}} & \bfseries 0.892 & 0.781 & \bfseries 0.855 & \bfseries 0.830 & 0.644 & \bfseries 0.713 & \bfseries 0.786\\
            & &\textit{cainet} & 0.827 & \bfseries 0.812 & 0.783 & 0.816 & \bfseries 0.681 & \bfseries 0.713 & 0.772\\
            & &\textit{fm4net} & 0.892 & 0.736 & 0.747 & 0.748 & 0.634 & 0.710 & 0.745\\

            \bottomrule
		\end{tabular}
	\end{minipage}
    \caption{\small{Quantitative results on PAD UFES 20 for the metadata impact on training analysis. Using the correlation coefficient computed between all metadata labels and the \textit{diagnostic} label we picked five and then labels with \textbf{h}ighest corresponding \textbf{c}orrelation coefficient (HC-5 and HC-10) and with the \textbf{l}owest \textbf{c}orrelation coefficients (LC-5 and LC-10) from Fig.~\ref{fig:corrcoeff}. We compare results for three fusion networks: \textit{vitatt} (ours, highlighted by the gray row), \textit{cainet} (Cai~\etal~\cite{cainet}), \textit{fm4net} (FusionM4Net \cite{fusionm4net}). The highest metric value for intra-class and average metrics has been marked in bold font.}}
    \label{fig:bestworst-results}
\end{figure}

\subsection{Interpretability}

We first explore the discriminative behavior of the trained \textit{vitatt} via a t-distributed Stochastic Neighbor Embedding Technique (t-SNE) based visualization of the generated latent space distribution. Fig.~\ref{fig:latent-space} shows the embedding vectors corresponding to all instances of the test set in PAD UFES 20 from the latent space both before and after the fusion stage. This allows us to see differences in model operation on single- and multi-modal embeddings.

The transformer-based architecture natively enables interpretability of its output through the attention mechanism. As noted by Chefer~\etal~\cite{te, tmme}, however, individual attention maps provide limited representation of the overall behavior of the model. We highlight \textit{vitatt}'s feature extraction performance in the image and patient metadata domain via the saliency maps provided by the TMME method \cite{tmme}.

\renewcommand{\x}{0.31}
\begin{figure}[!h]
    \small
	\centering
	\begin{minipage}{0.02\linewidth}
		\centering
		\hfill
	\end{minipage}
	\begin{minipage}{0.62\linewidth}
		\centering
		Correctly classified lesions
	\end{minipage}
	\begin{minipage}{\x\linewidth}
		\centering
		Misclassified lesions
	\end{minipage}
	\\
	\begin{minipage}{0.02\linewidth}
		\centering
		\rotatebox[origin=c]{90}{\footnotesize NEV}
	\end{minipage}
	\begin{minipage}{\x\linewidth}
		\centering
		\includegraphics[width=\linewidth]{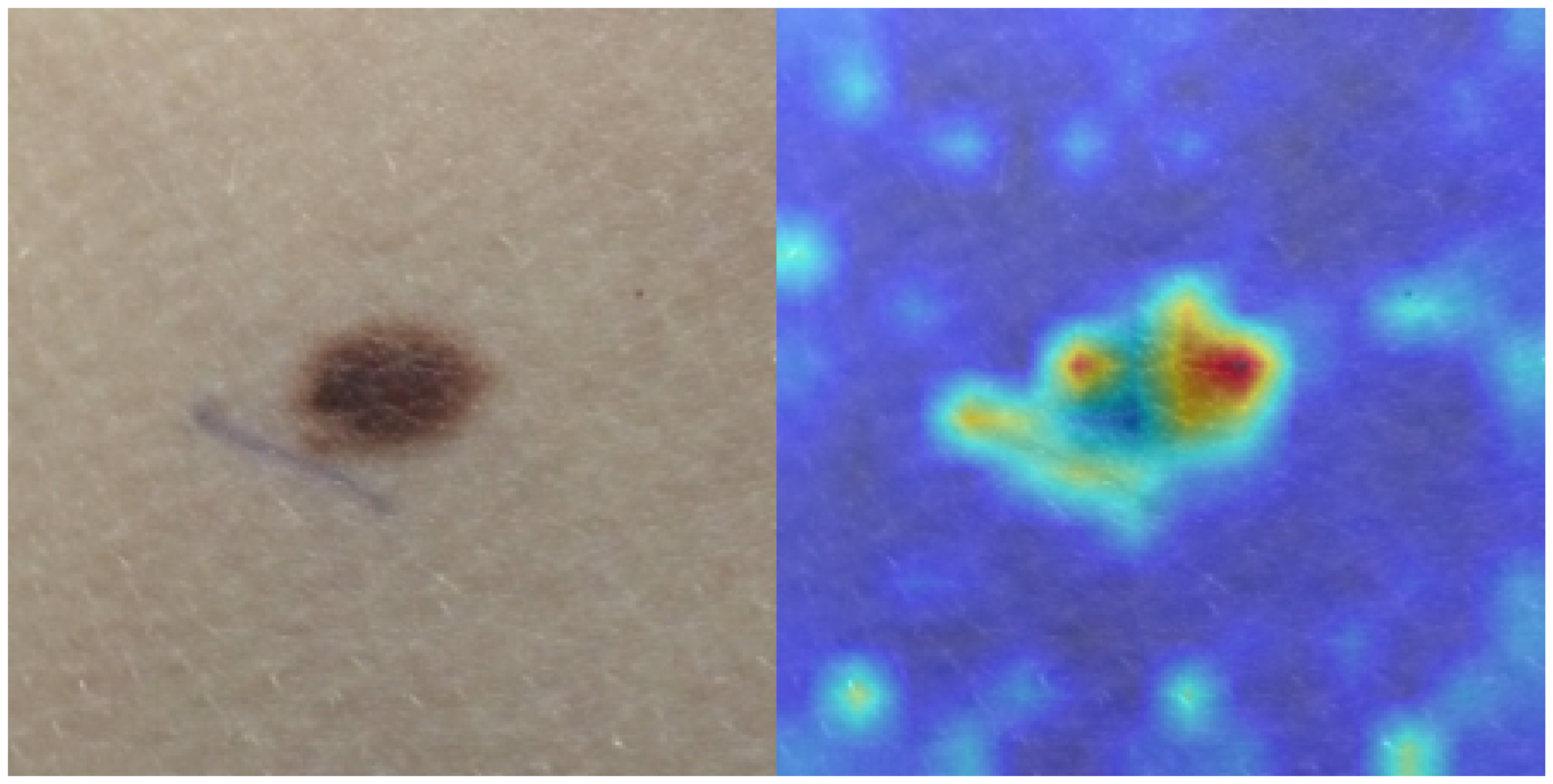}
	\end{minipage}
	\begin{minipage}{\x\linewidth}
		\centering
		\includegraphics[width=\linewidth]{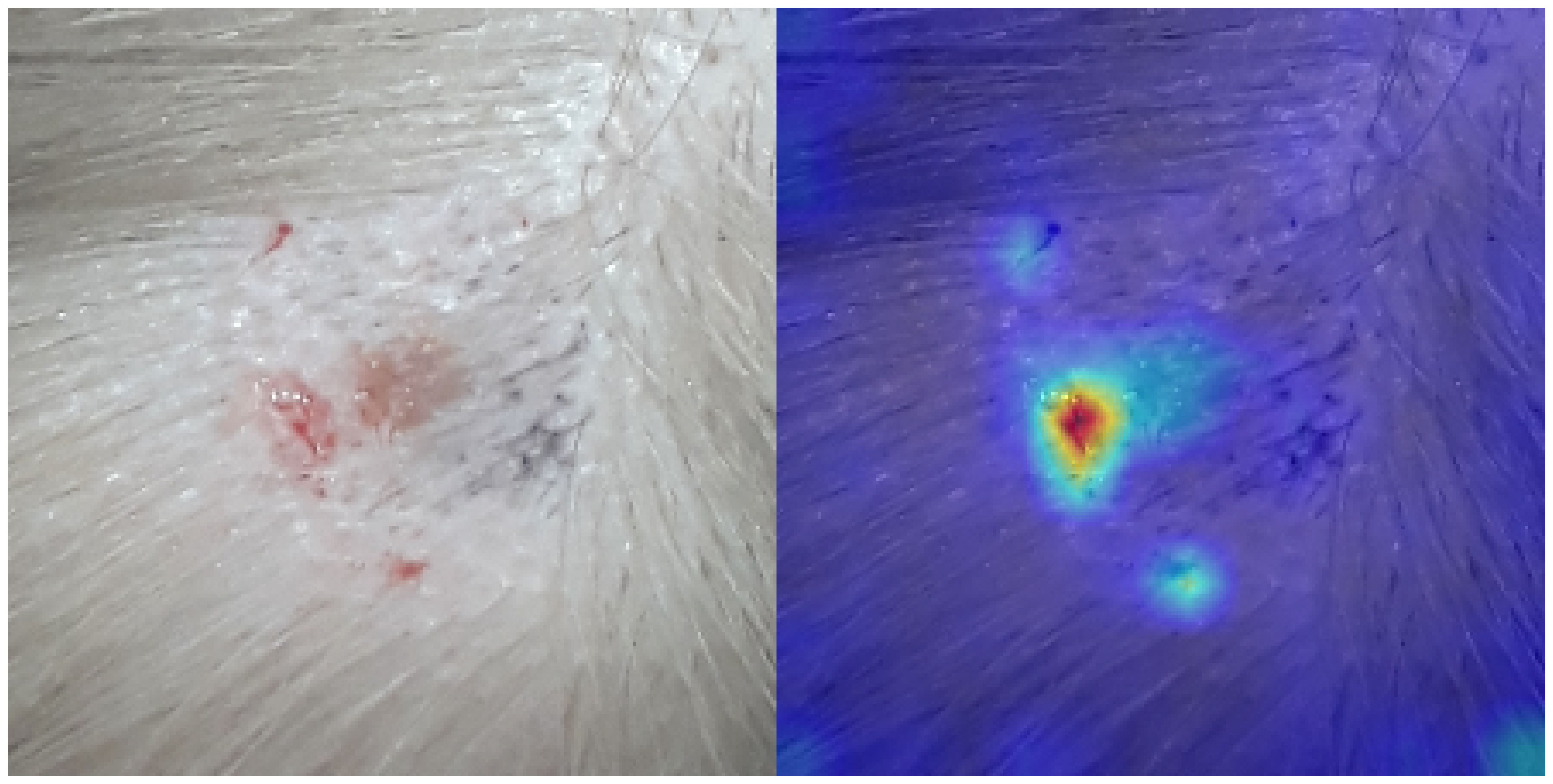}
	\end{minipage}
	\begin{minipage}{\x\linewidth}
		\centering
		\includegraphics[width=\linewidth]{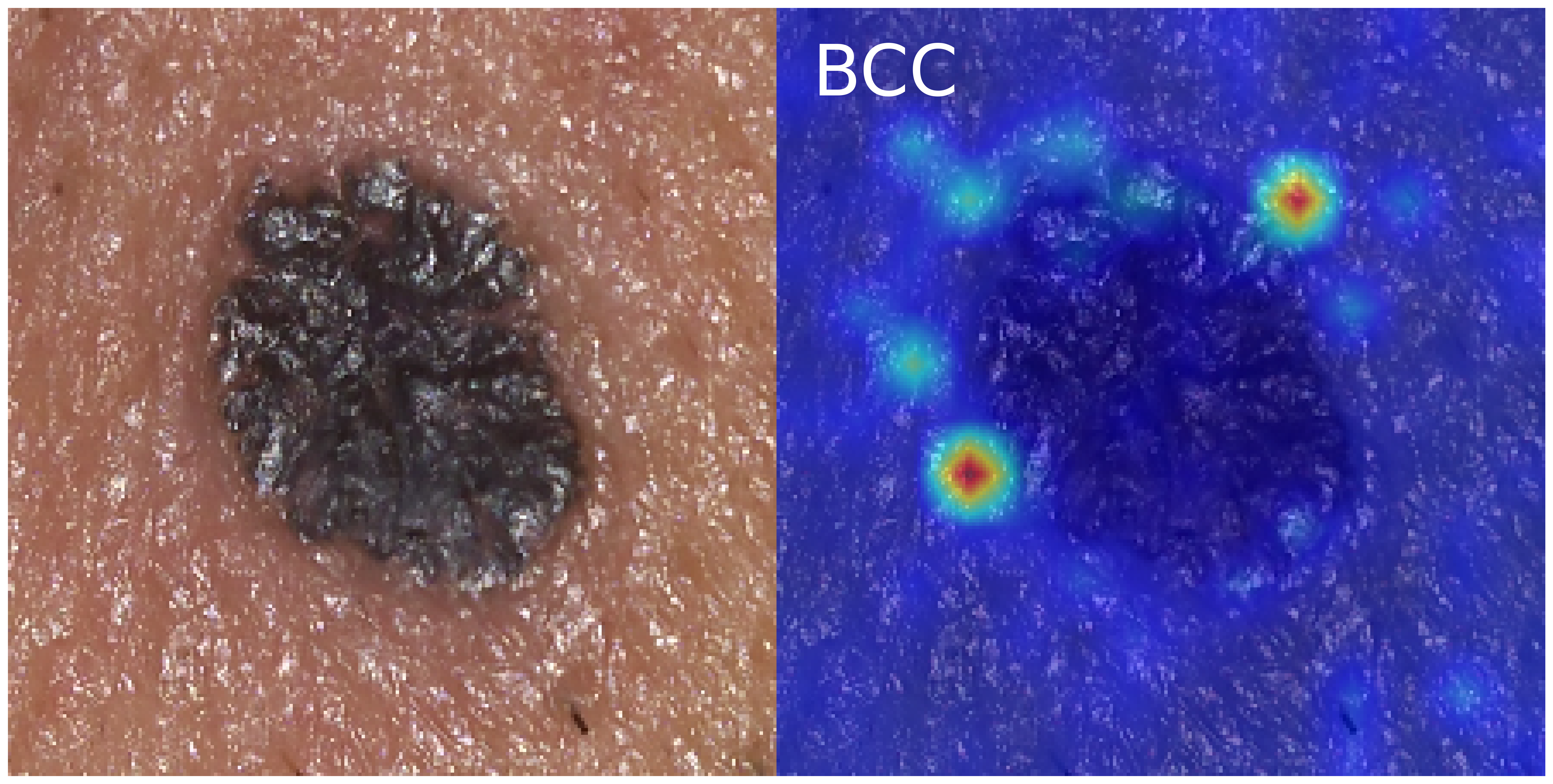}
	\end{minipage}
	\\
	\begin{minipage}{0.02\linewidth}
		\centering
		\rotatebox[origin=c]{90}{\footnotesize BCC}
	\end{minipage}
	\begin{minipage}{\x\linewidth}
		\centering
		\includegraphics[width=\linewidth]{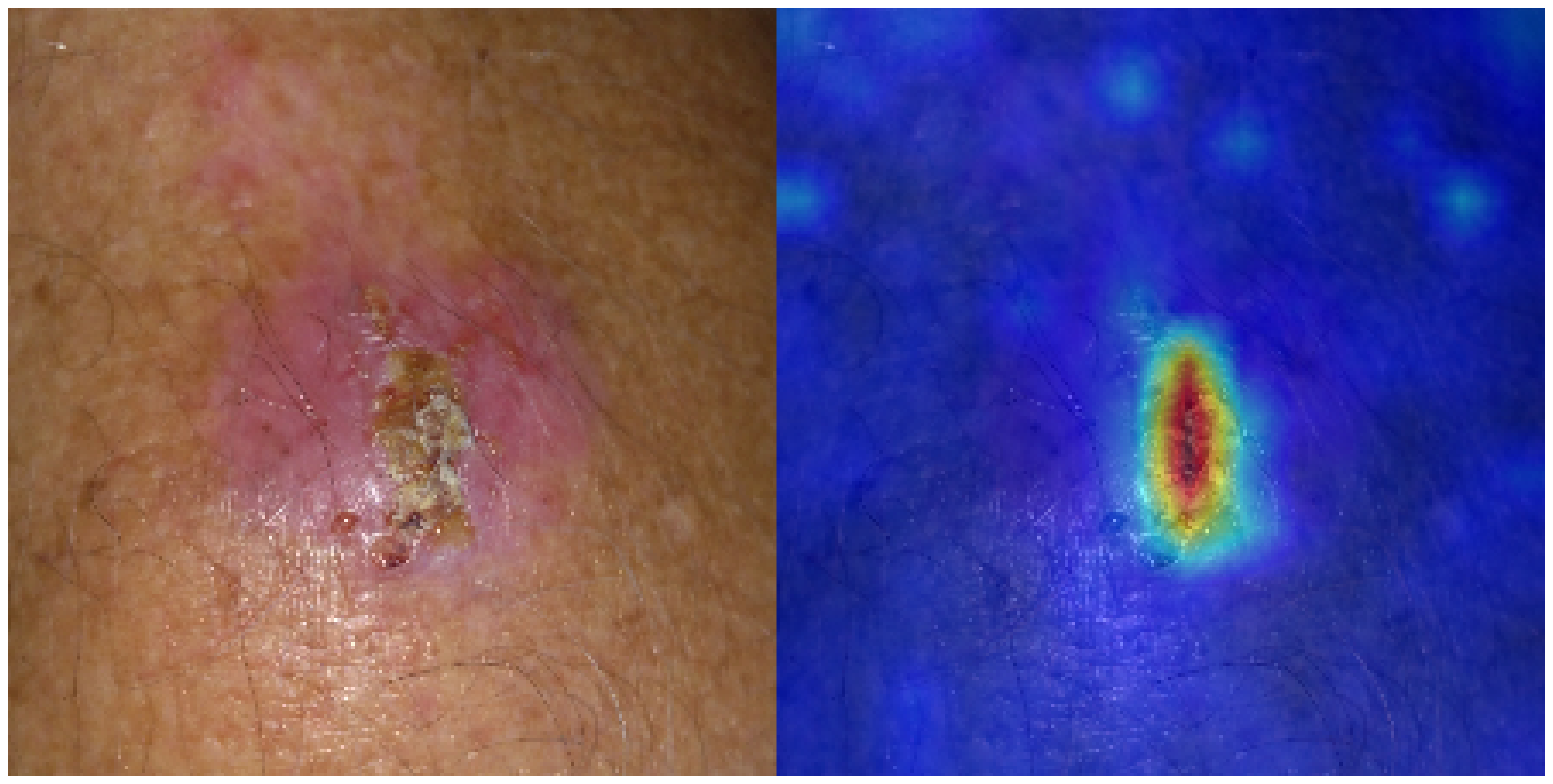}
	\end{minipage}
	\begin{minipage}{\x\linewidth}
		\centering
		\includegraphics[width=\linewidth]{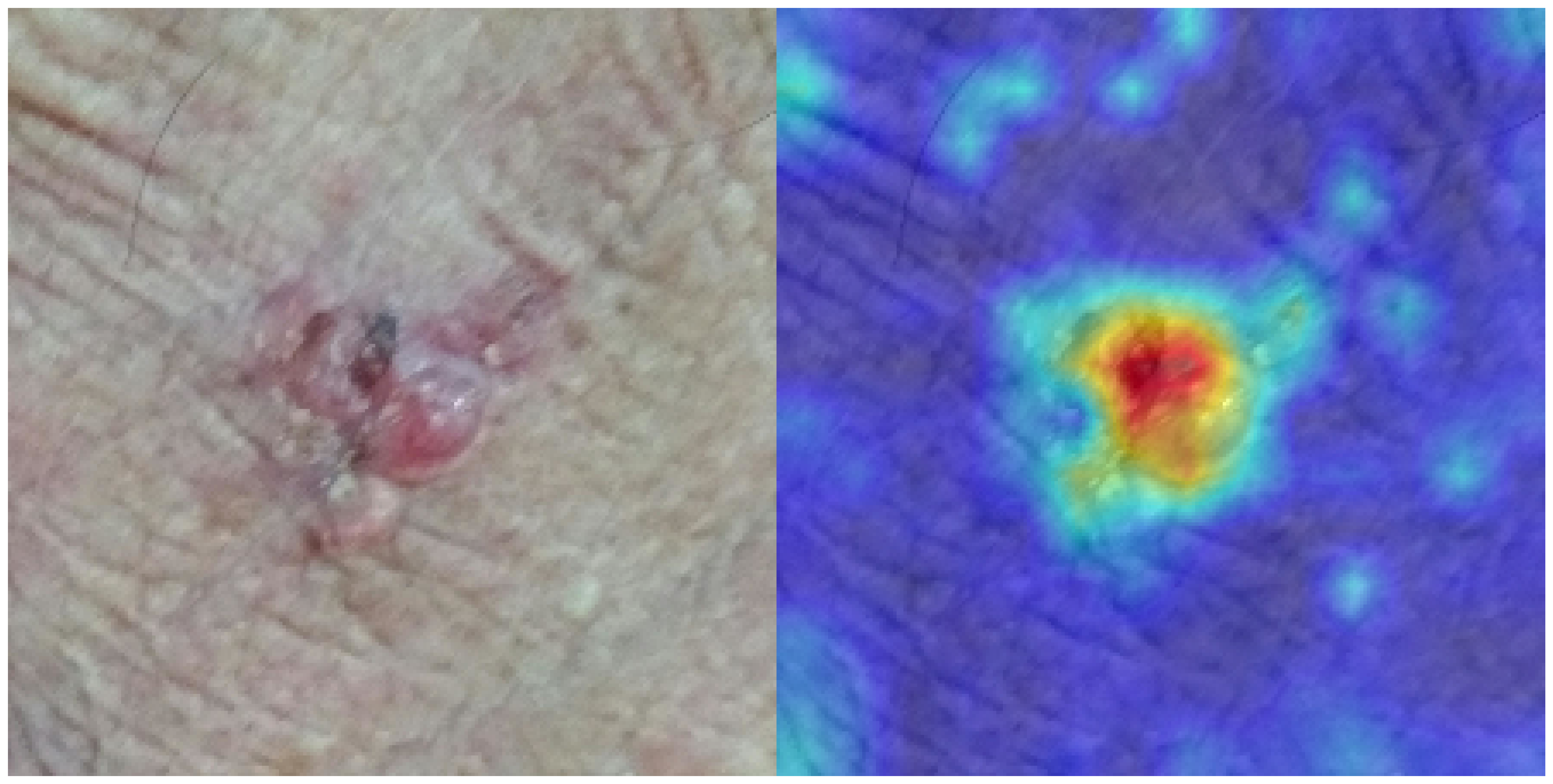}
	\end{minipage}
	\begin{minipage}{\x\linewidth}
		\centering
		\includegraphics[width=\linewidth]{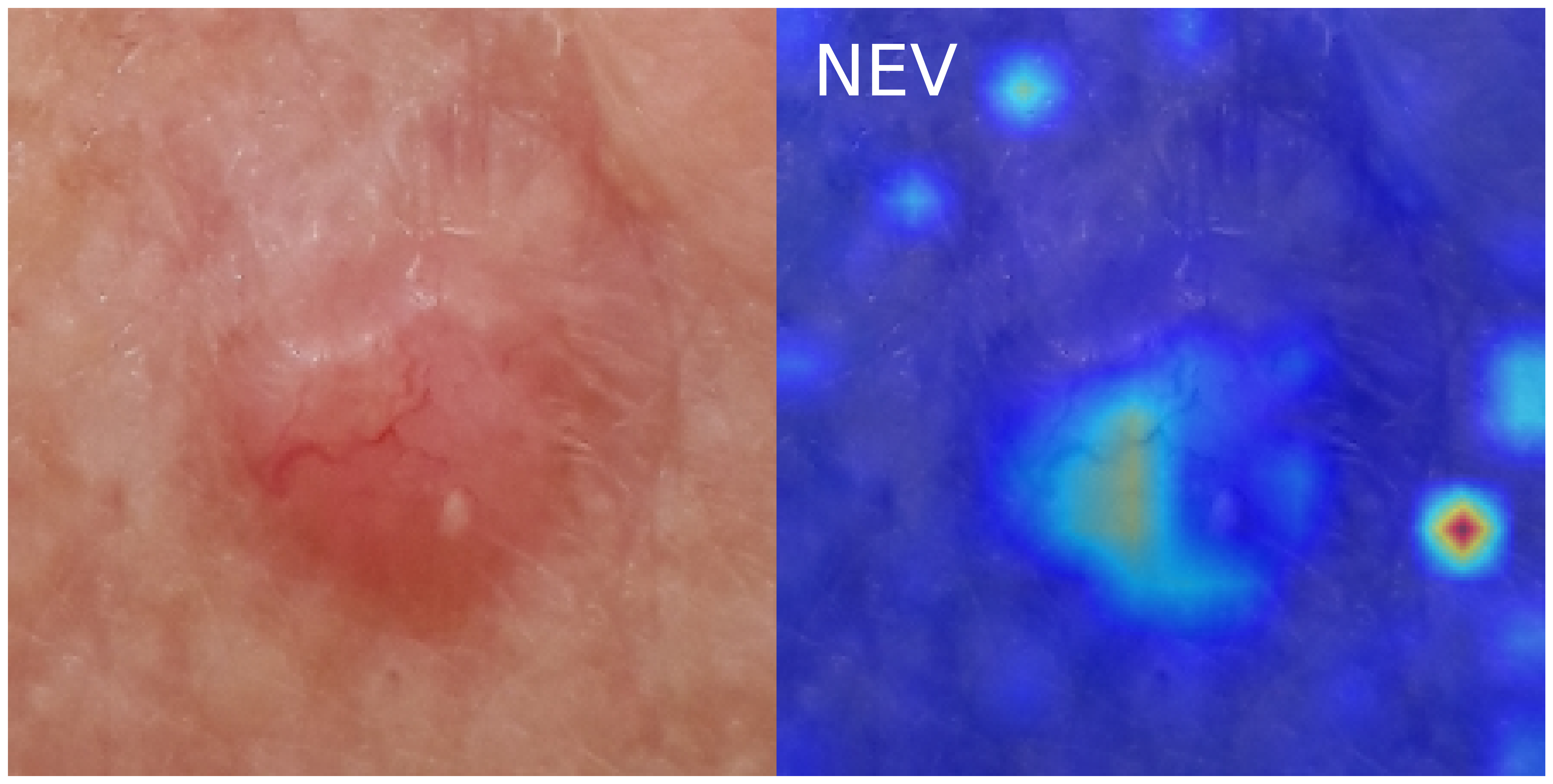}
	\end{minipage}
	\\
	\begin{minipage}{0.02\linewidth}
		\centering
		\rotatebox[origin=c]{90}{\footnotesize MEL}
	\end{minipage}
	\begin{minipage}{\x\linewidth}
		\centering
		\includegraphics[width=\linewidth]{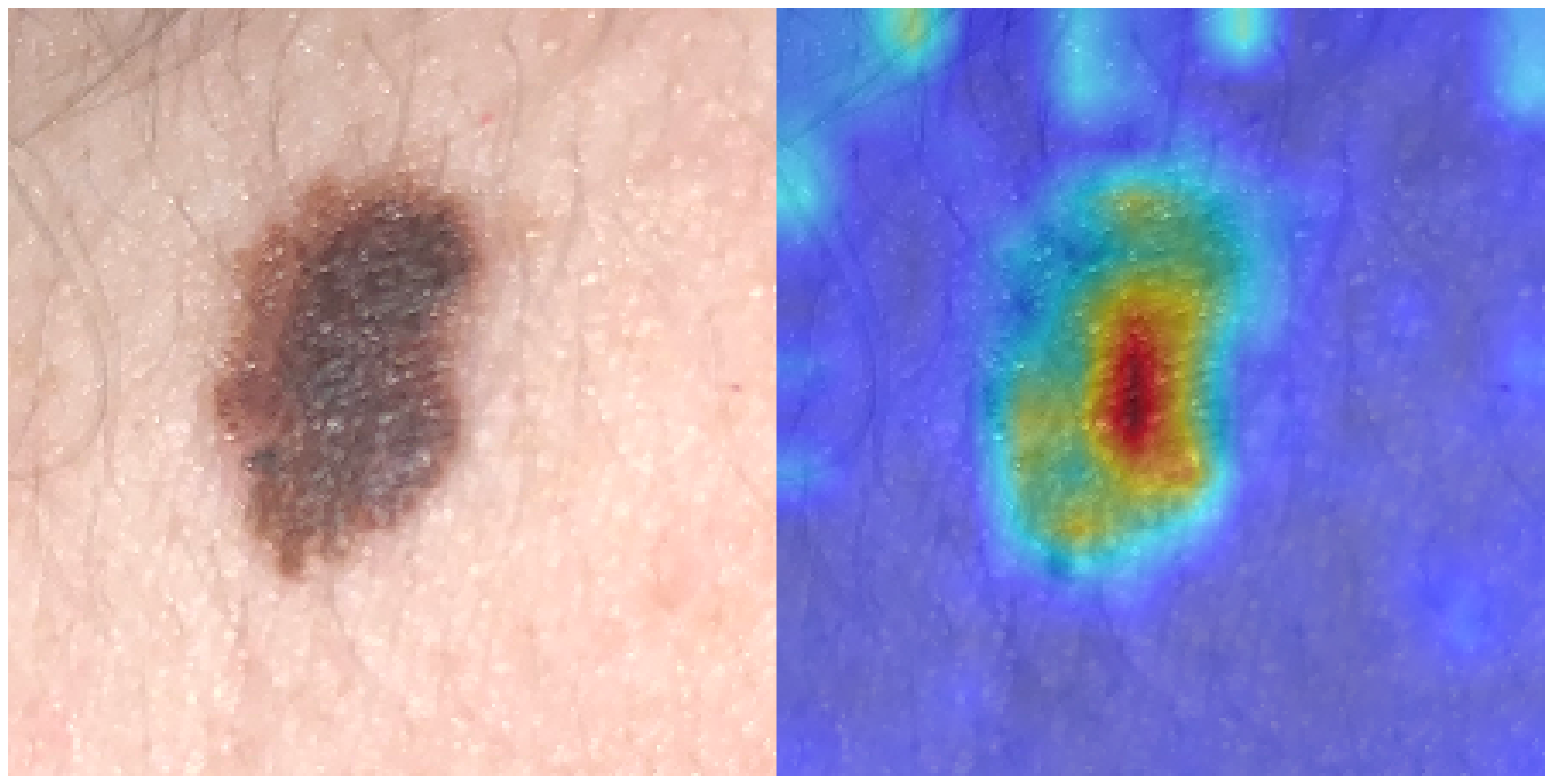}
	\end{minipage}
	\begin{minipage}{\x\linewidth}
		\centering
		\includegraphics[width=\linewidth]{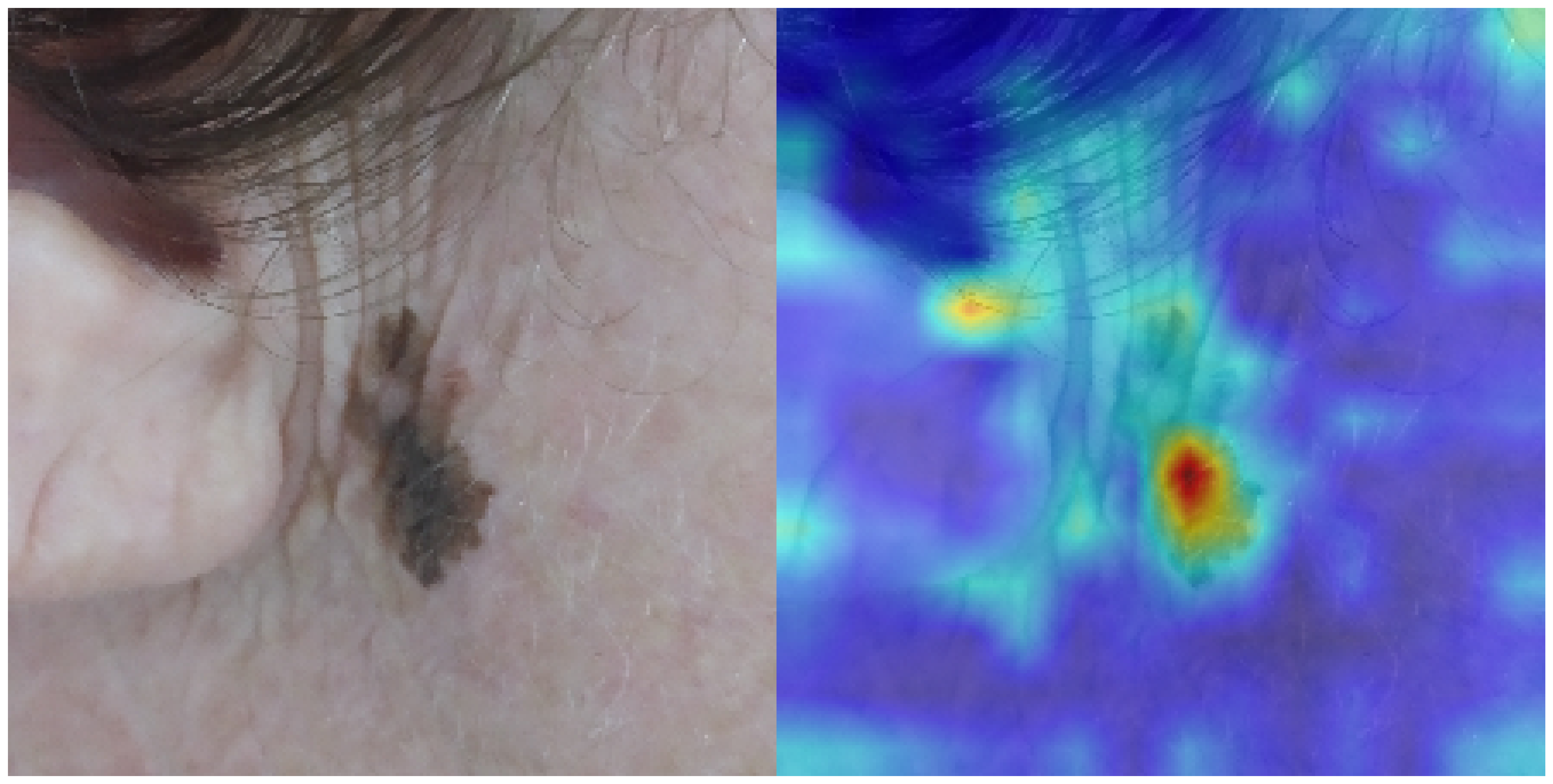}
	\end{minipage}
	\begin{minipage}{\x\linewidth}
		\centering
		\includegraphics[width=\linewidth]{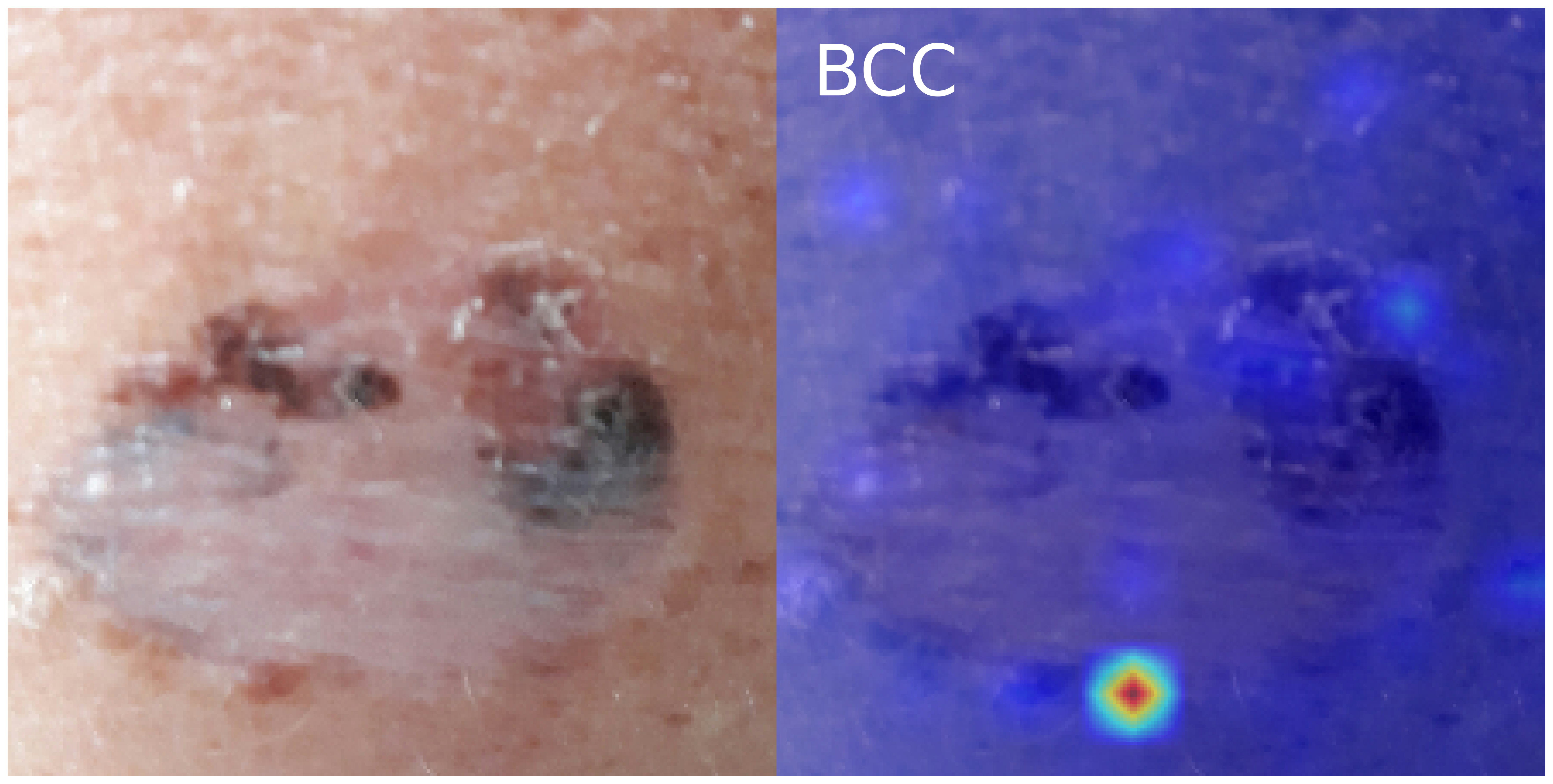}
	\end{minipage}
	\\
	\begin{minipage}{0.02\linewidth}
		\centering
		\rotatebox[origin=c]{90}{\footnotesize  ACK}
	\end{minipage}
	\begin{minipage}{\x\linewidth}
		\centering
		\includegraphics[width=\linewidth]{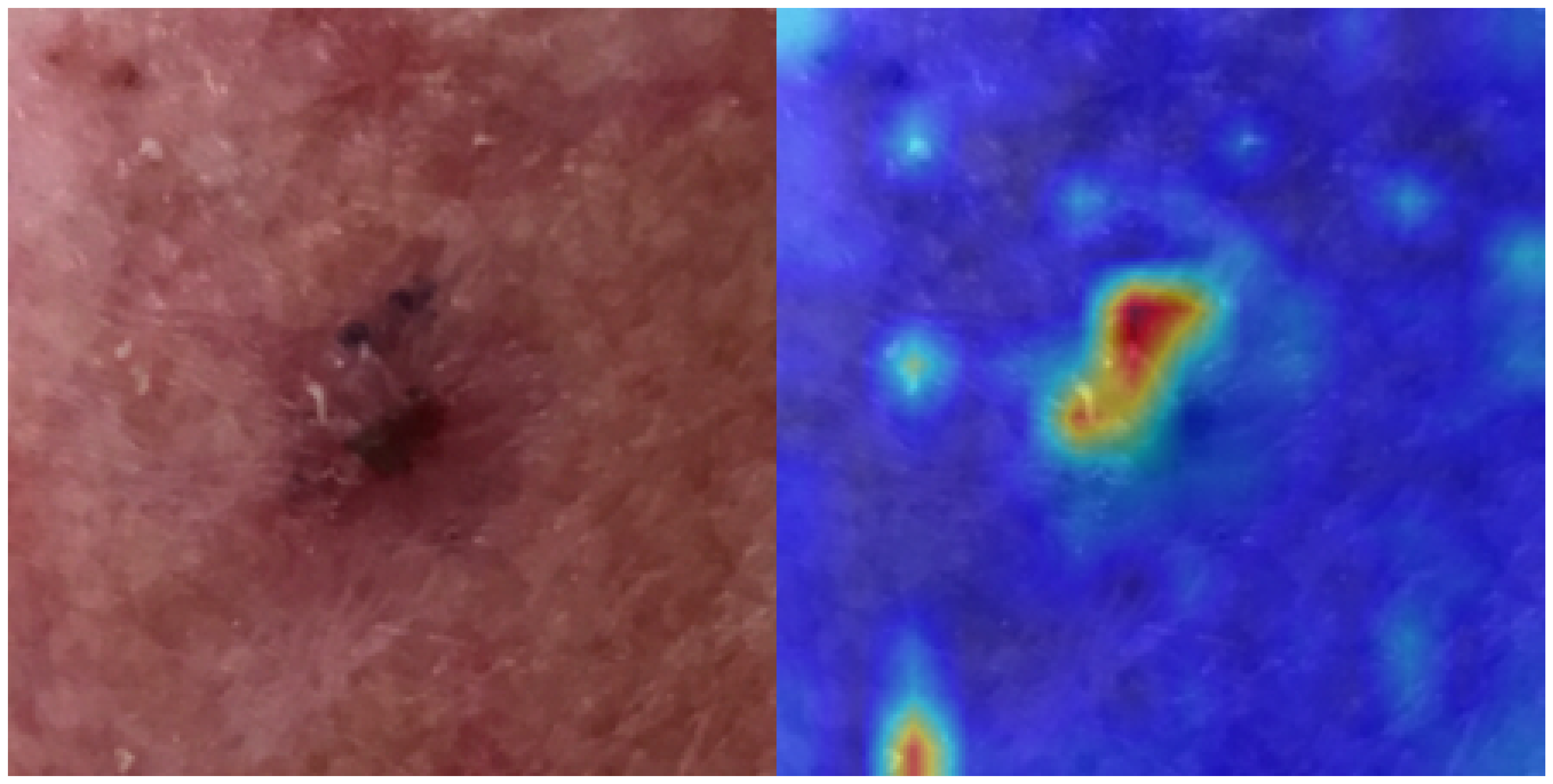}
	\end{minipage}
	\begin{minipage}{\x\linewidth}
		\centering
		\includegraphics[width=\linewidth]{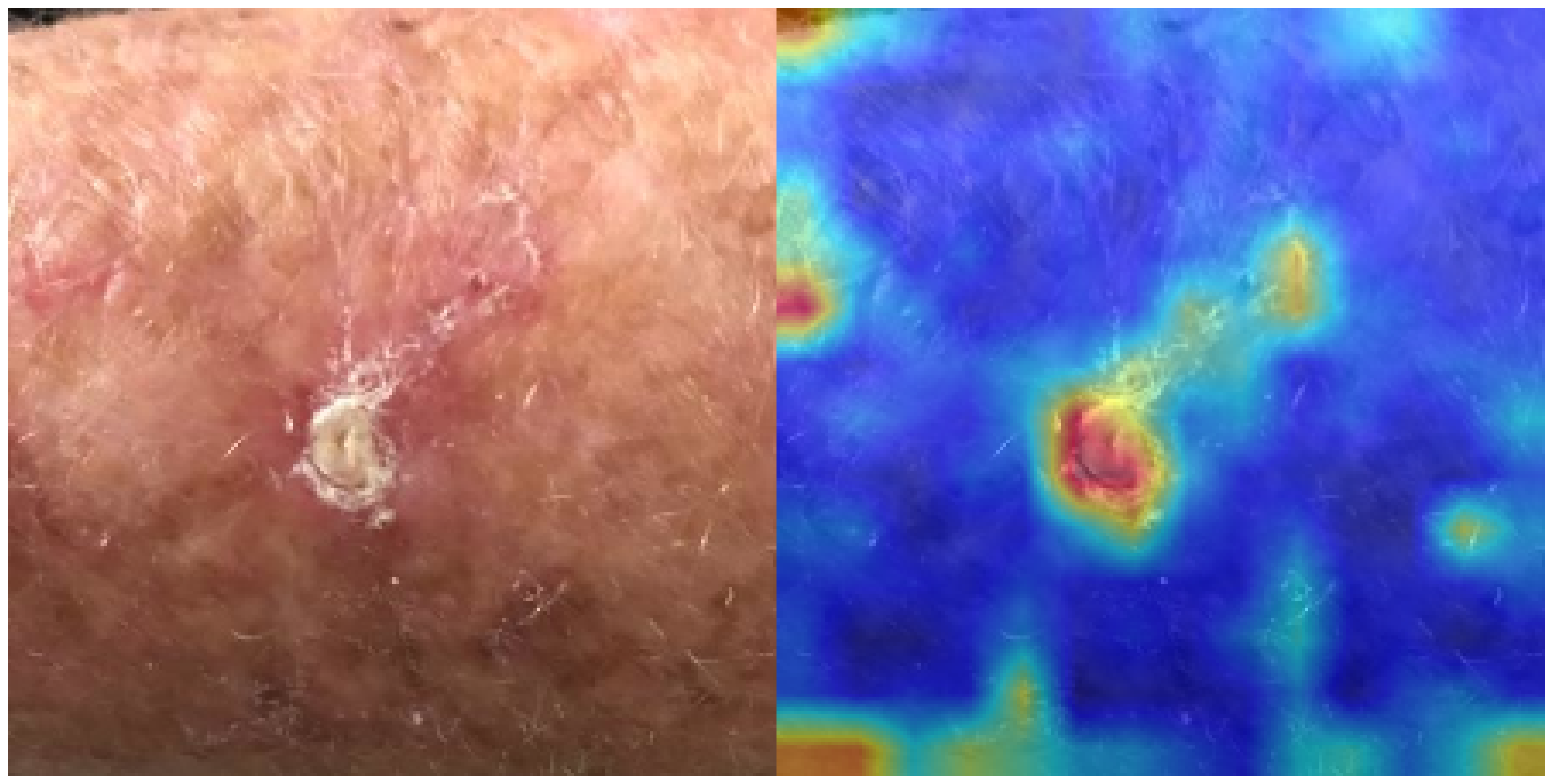}
	\end{minipage}
	\begin{minipage}{\x\linewidth}
		\centering
		\includegraphics[width=\linewidth]{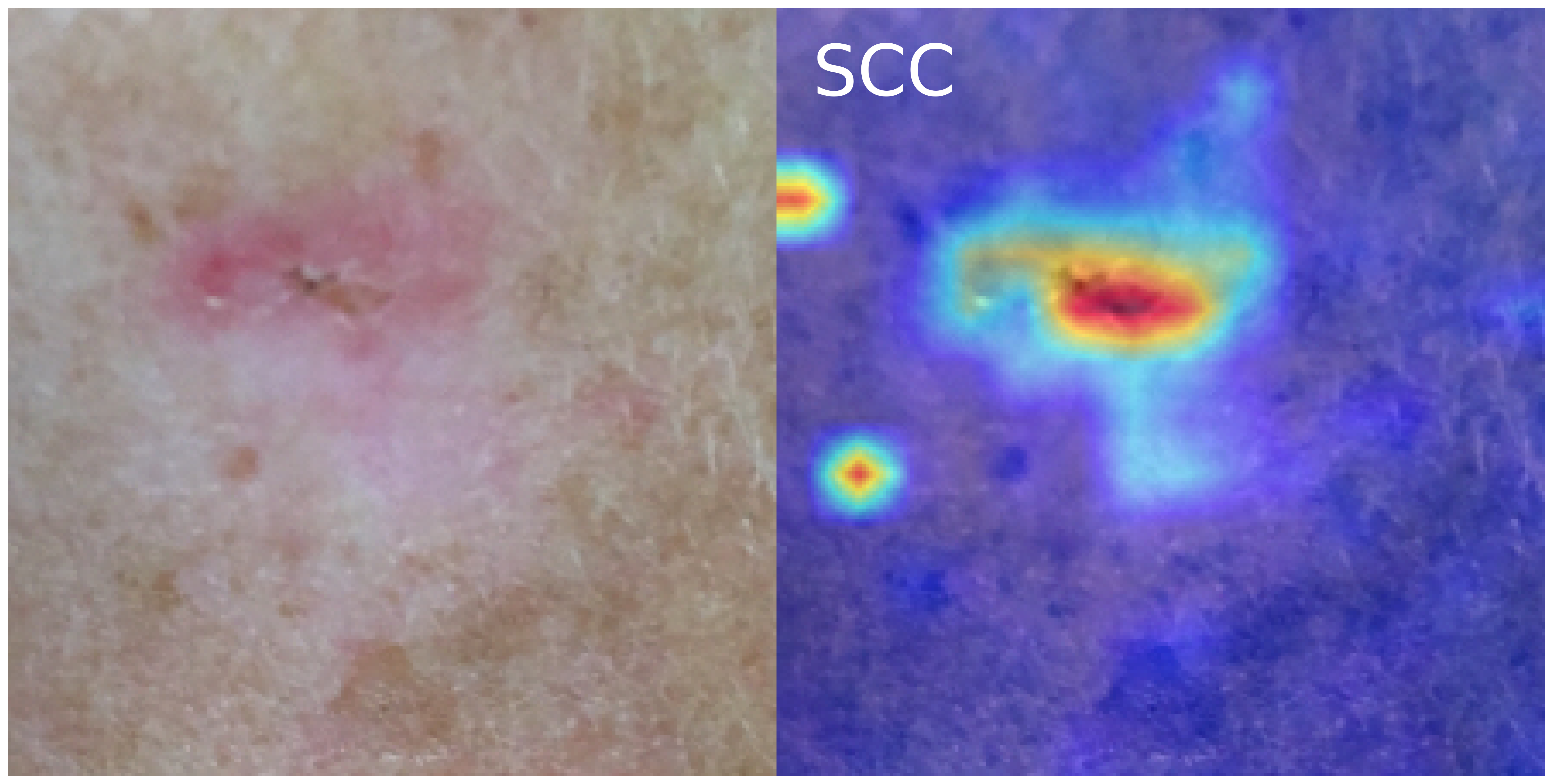}
	\end{minipage}
	\\
	\begin{minipage}{0.02\linewidth}
		\centering
		\rotatebox[origin=c]{90}{\footnotesize SCC}
	\end{minipage}
	\begin{minipage}{\x\linewidth}
		\centering
		\includegraphics[width=\linewidth]{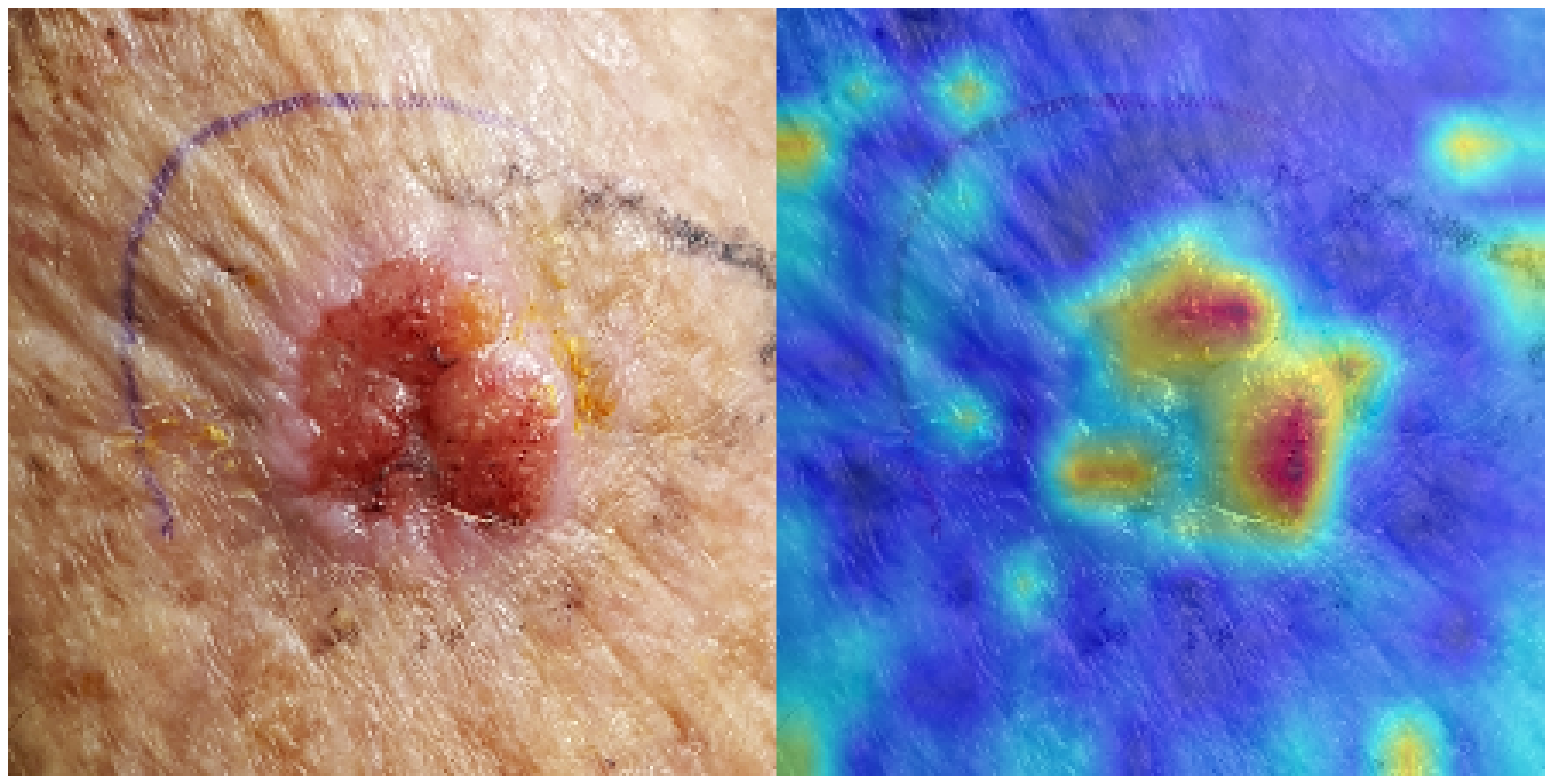}
	\end{minipage}
	\begin{minipage}{\x\linewidth}
		\centering
		\includegraphics[width=\linewidth]{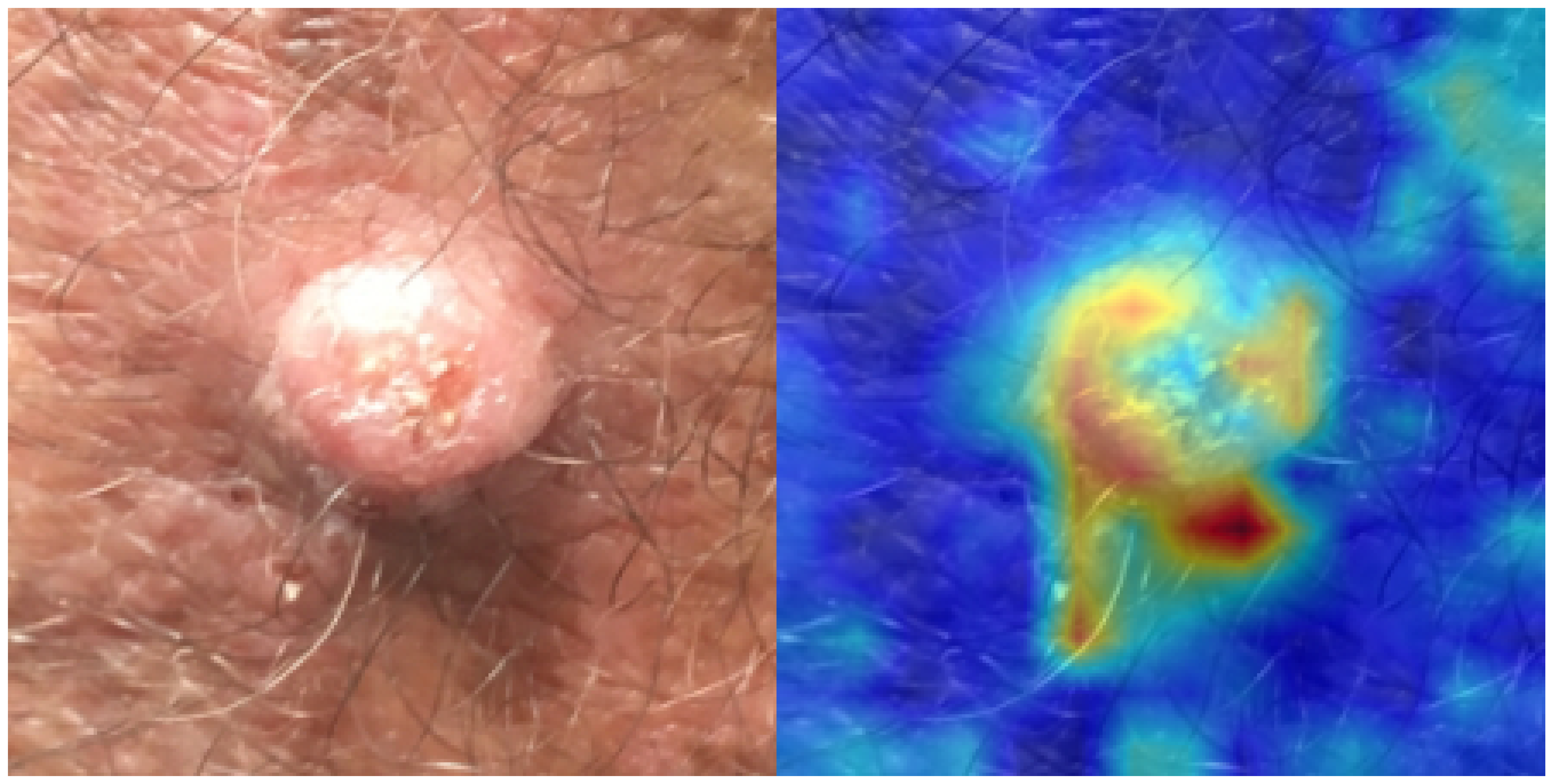}
	\end{minipage}
	\begin{minipage}{\x\linewidth}
		\centering
		\includegraphics[width=\linewidth]{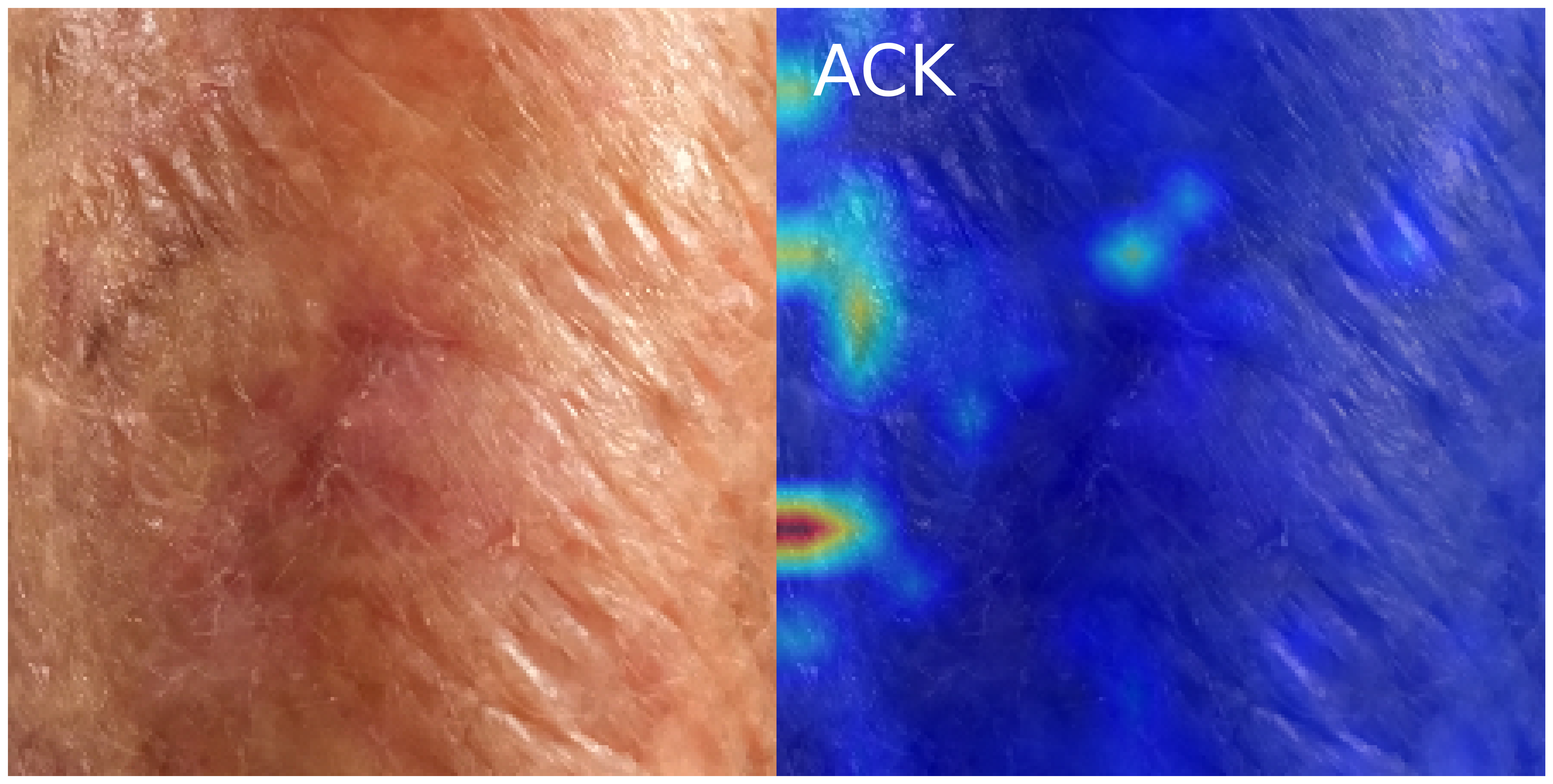}
	\end{minipage}
	\\
	\begin{minipage}{0.02\linewidth}
		\centering
		\rotatebox[origin=c]{90}{\footnotesize SEK}
	\end{minipage}
	\begin{minipage}{\x\linewidth}
		\centering
		\includegraphics[width=\linewidth]{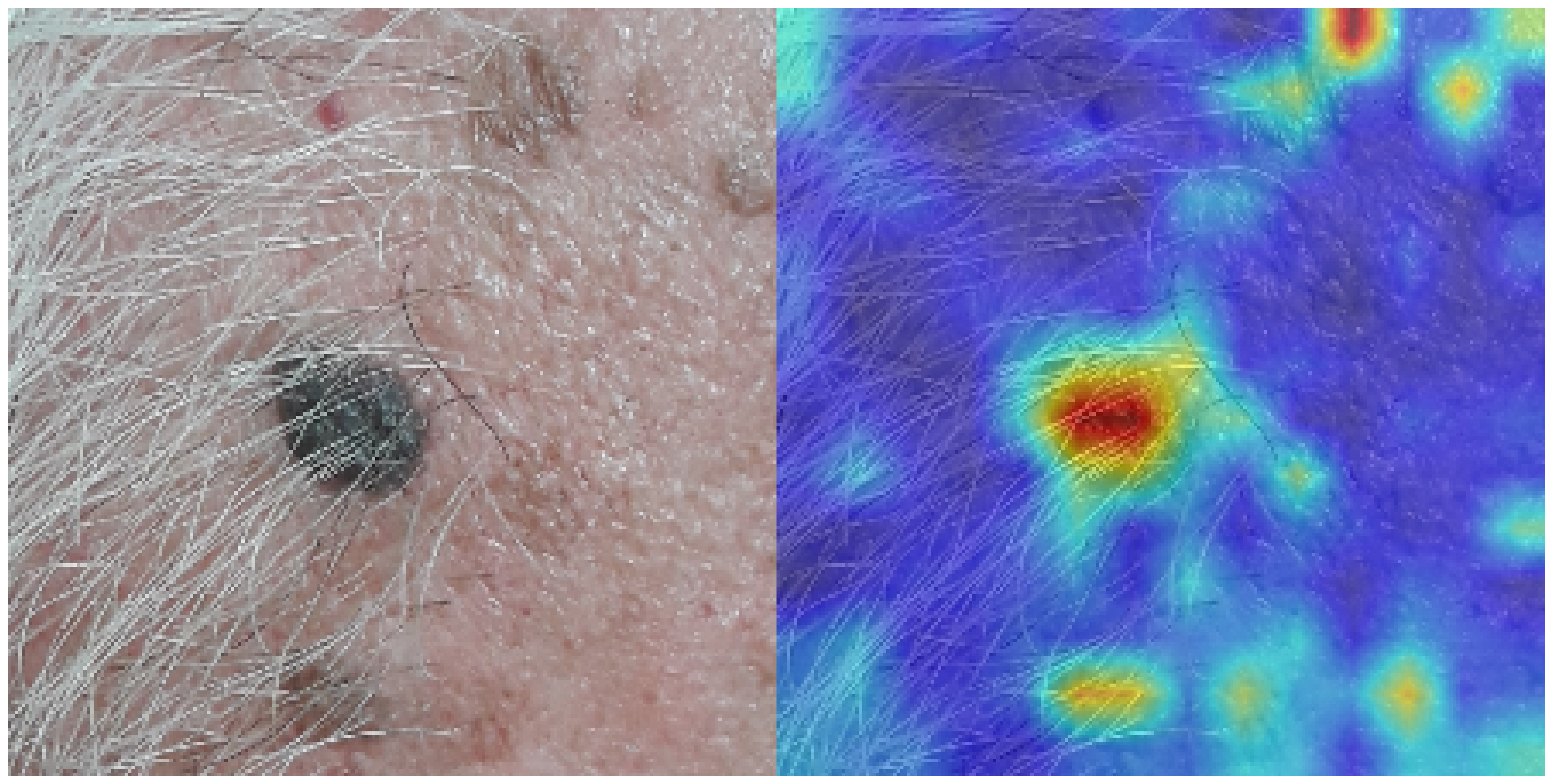}
	\end{minipage}
	\begin{minipage}{\x\linewidth}
		\centering
		\includegraphics[width=\linewidth]{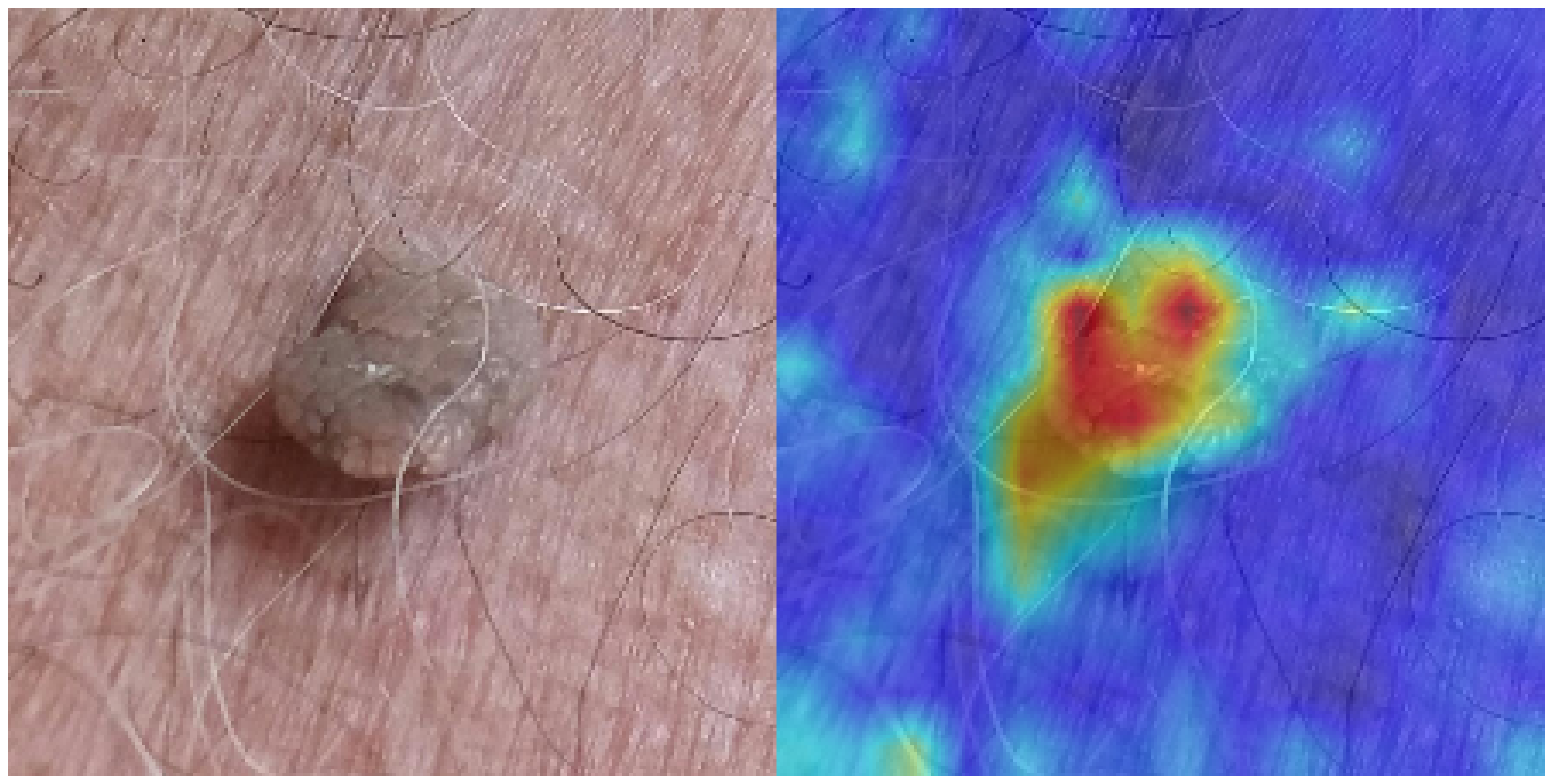}
	\end{minipage}
	\begin{minipage}{\x\linewidth}
		\centering
		\includegraphics[width=\linewidth]{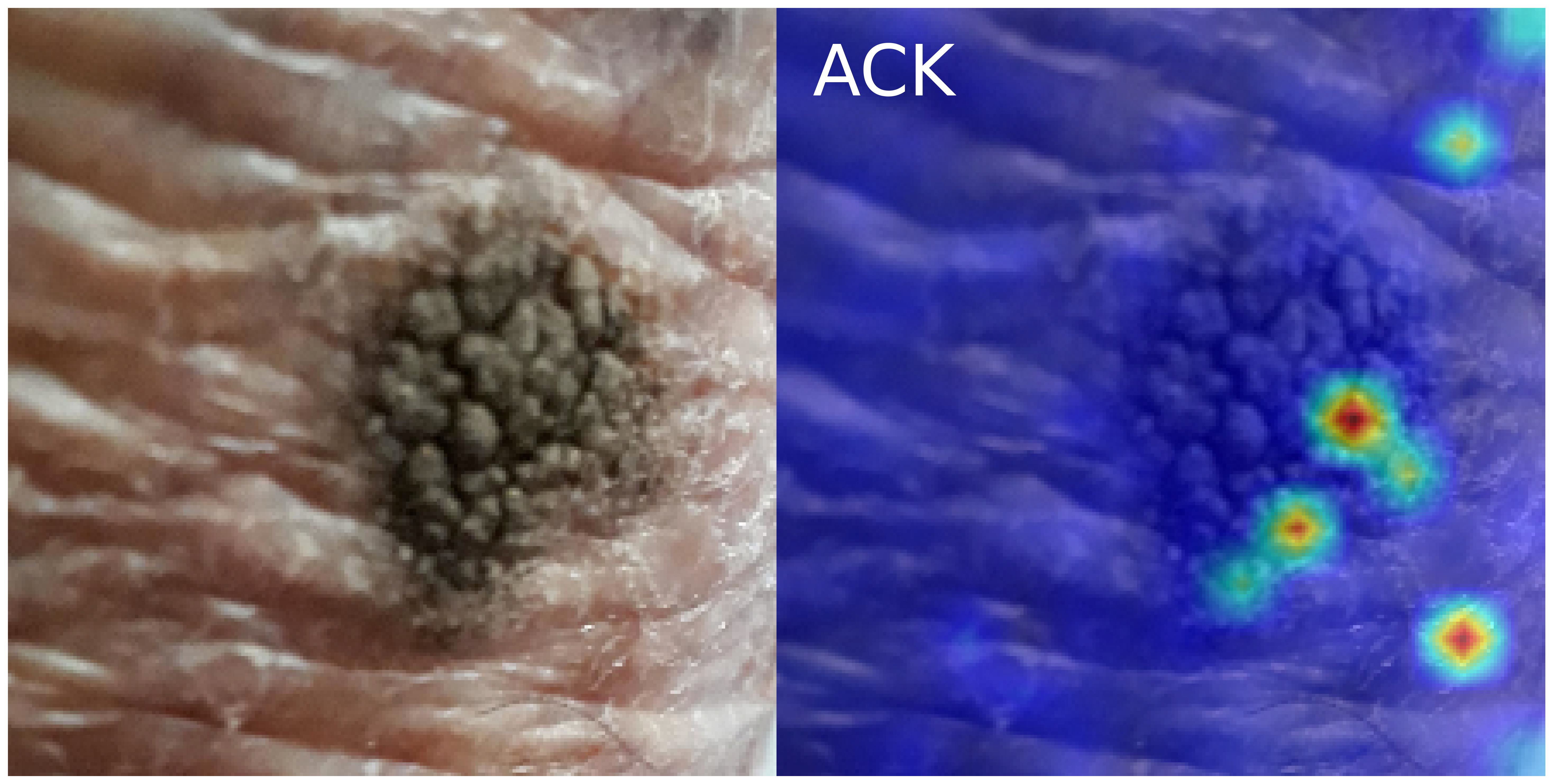}
	\end{minipage}
	\caption{\small{Results with corresponding TMME Saliency Maps for each class (\textbf{rows}).  Since the images have been obtained with a smartphone, they differ in several aspects from one another: brightness, contrast, vibrance. The saliency maps contain regions marked by the trained network as of \enquote{high importance} for the final classification in red and of \enquote{low importance} in blue. The first two columns show correctly classified images, while the third column shows incorrectly classified lesions with the corresponding saliency map generated for the incorrect label. The wrongly classified label has been added at top-left of each map.}}
	\label{fig:tmme-image}
\end{figure}

\begin{figure*}[t]
	\centering
    \begin{minipage}{0.65\linewidth}
        \includegraphics[width=\linewidth]{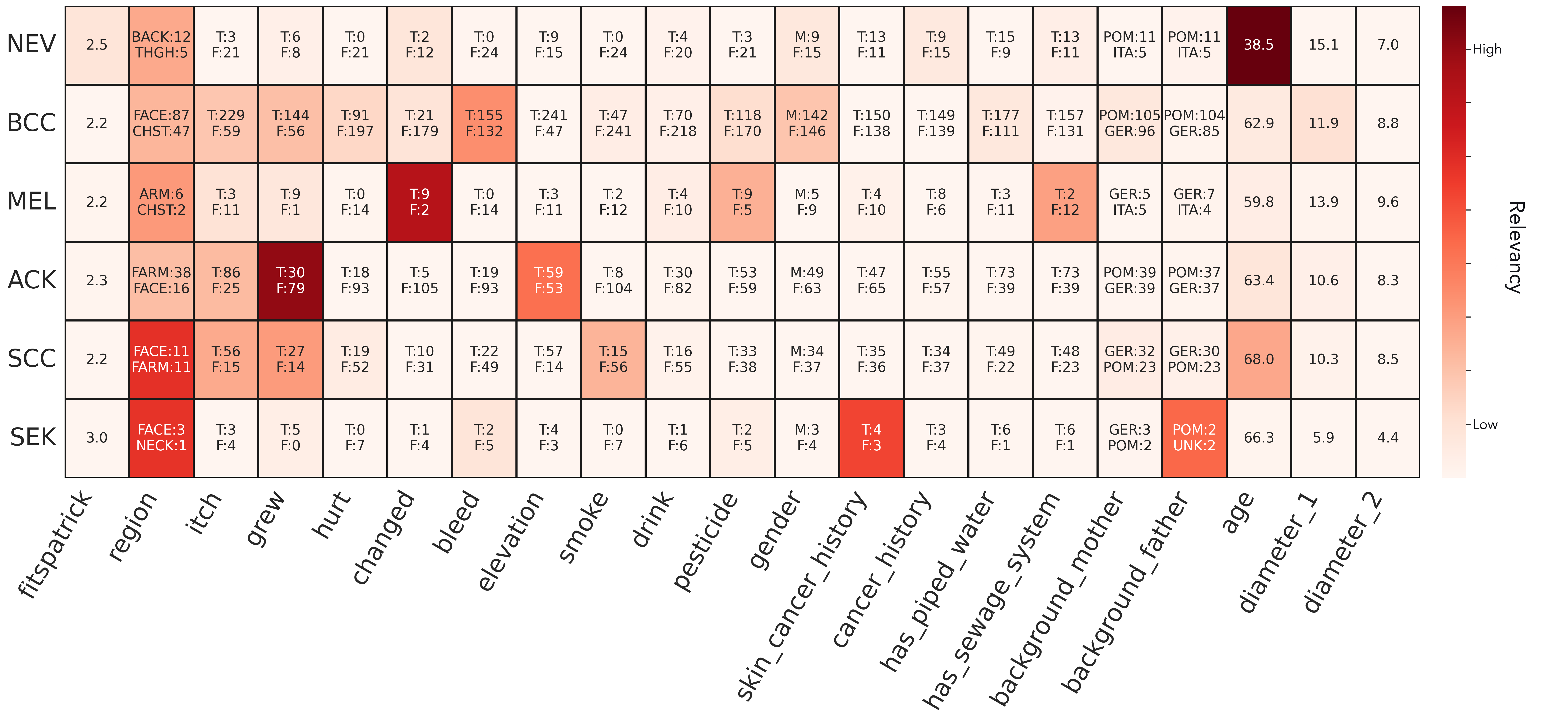}
    \end{minipage}
    \begin{minipage}{0.34\linewidth}
        \caption{\small{TMME for the Metadata Branch with relevancy coefficients (higher is better) for all (class, metadata label) pairs. We averaged the computed TMME relevancy maps for all instances in the test set over each disease class (\textbf{rows}). Every cell contains additional information about the distribution of metadata values as follows: binary metadata information is shown as T(rue), F(alse) with their respective number of instances, the \textit{region}, \textit{background\_father} and \textit{background\_mother} labels show the two most common values inside the each disease class with respective instance count, and the \textit{fitspatrick}, \textit{age}, \textit{diameter\_1,} \textit{diameter\_2} label cell contain the average values.}}
        \label{fig:tmme-metadata}
    \end{minipage}
\end{figure*}

Fig. \ref{fig:tmme-image} contains three images for each of the six classes represented in PAD UFES 20, each with its corresponding saliency map. The images are all correctly classified by the trained \textit{vitatt} model. Fig.~\ref{fig:tmme-metadata} contains separate average relevancy maps of the patient metadata for all six classes represented in PAD UFES 20 and the respective metadata. The value of the relevancy score of a particular disease-metadata pair corresponds to the impact of that pair on the final classification output (equivalent to the high saliency areas (red) in Fig. \ref{fig:tmme-image}).

\subsection{Implementation details}
To counter-act the dataset imbalance for both datasets, we employed weighted cross-entropy loss coupled with weighted random sampling of the datasets. The loss and sampler functions were defined via their Pytorch \textit{WeightedCrossEntropy} and \textit{WeightedRandomSampler} implementations.

We trained \textit{vitatt} for 200 epochs on both datasets, with a batch size of 16, and a custom learning rate \cite{lrperlayer} defined as $\text{\textit{lr}} = 3\cdot10^{-5}$ if layer is in the transformer encoder and $\text{\textit{lr}} = 10^{-4}$ otherwise. The optimizer was \textit{Adam}, and we experimentally concluded that using no weight decay leads to better results.

Each model was trained five times and only the mean metric values were plotted in Figs.~\ref{fig:main-results} and \ref{fig:bestworst-results} to declutter the results.

\section{Discussion} \label{section:discussion}

\subsection{Technical Aspects}
The results in Fig.~\ref{fig:main-results} show that the \textit{vitatt} model outperforms all the state-of-the-art methods on both the image-poor high-quality metadata dataset (PAD UFES 20) and the high-quality image few-metadata dataset (ISIC 2019). As expected, both \textit{vitatt} and \textit{cainet} lead to higher quantitative metrics compared to the two-stage \textit{fm4net}, mainly thanks to the nature of the architecture allowing for the classification scores to be computed on the multi-modal fused latent features. Compared to \textit{cainet}, \textit{vitatt} proves to be more effective in extracting relevant multi-modal features from both input domains thanks to the ability of \textit{vitatt} to perform all pairs of attention operations in a single step and branch: image-to-image, image-to-\textbf{class}, image-to-metadata, metadata-to-metadata, and metadata-to-\textbf{class}. On PAD UFES 20, for classes with many training samples (ACK, BCC), \textit{vitatt} has at least 6\% higher accuracy and AUC. For the NEV class, which has fewer samples than BCC, we are still better than the other fusion models. For sparse classes like SEK and SCC, no model of the six analyzed produces any significantly positive results. Their high intra-class variance (see Fig.~\ref{fig:tmme-image}) and the low number of training samples impact the performance of all models. For ISIC 2019, \textit{effnet} has a slightly better performance per class than any other network (see melanoma precision in Fig.~\ref{fig:main-results}) but on average lags behind \textit{vitatt} in all metrics.

Regarding the metadata study, Fig.~\ref{fig:bestworst-results} contains the quantitative metrics computed for the three multi-modal classifiers. \textit{cainet} performs better than our \textit{vitatt} in the LC-5 case, as it is deeper and contains two branches with cross-modal attention operation instead of only one self-attention layer. However, as we increase the number of metadata points (LC-10) and their computed correlation (HC-5 and HC-10) we outperform \textit{cainet} (e.g. AUC), as \textit{cainet} tends to overfit on the training set due to having three times as many parameters as \textit{vitatt}.

The latent space visualization in Fig.~\ref{fig:latent-space} sheds more light onto the discriminative behavior improvement triggered by the fusion layer. Before the fusion layer, the \textbf{class} token only attends on the image data. The low amount of image data in PAD UFES 20 used for training coupled with the data-hungry nature of the transformer encoder architecture significantly contributes to the overlap of the different latent class clusters. The attention-based fusion layer enables the \textbf{class} token to also attend on the embedded patient metadata, which in turn leads to a better separation of the clusters.

In Fig.~\ref{fig:tmme-image} we find the associated TMME relevancy maps for three correctly classified images of each class from PAD UFES 20. Intuitively, the model is capable of localizing and segmenting the lesion quite well, however, due to the imbalanced nature of the dataset, classes with more data (like BCC or NEV) generate better visualizations with less noise than lower-represented ones. The TMME maps also take in account the fusion attention layer, therefore, the relevancy map has the metadata attention information embedded into it. Good performance by \textit{vitatt} can also be seen when hair is present in the image, as the model is able to differentiate the lesion from its neighborhood. We also noticed that the pen-markings made by the dermatologist on the patient's skin are of interest for the model. Due to the small size of the dataset, the trained \textit{vitatt} is not fully capable of ignoring these markers.

Finally, Fig.~\ref{fig:tmme-metadata} shows learned correlations between metadata and the respective disease representation. In combination with Fig.~\ref{fig:tmme-image}, this figure highlights the model's interpretation of each skin disease based on the $21$ available metadata factors. This information can be then used to identify if clinically relevant diagnostic information is learned by the model.

\subsection{Medical Aspects}

Insights provided by the saliency and relevancy maps obtained with TMME in Figs.~\ref{fig:tmme-image} and \ref{fig:tmme-metadata} help the dermatologist identify failure points in the trained representation of the disease or correct reasoning for the classification (determining metadata factors for the confirmation of certain diseases in Fig.~\ref{fig:tmme-metadata}). For instance, non-medical cues like markers are deemed by the model \enquote{relevant} to the classification but do not lead to a misclassification (e.g. Fig.~\ref{fig:tmme-image} NEV first column). On the other hand, for a misclassified case, a dermatologist will use the provided saliency map to dismiss the false negative due to lacking diagnostic relevancy of the highlighted area (e.g. Fig.~\ref{fig:tmme-image} MEL third column).

The localization performance of the model on dense classes (NEV, BCC) is higher than for sparse classes (ACK, SEK), seen in Fig.~\ref{fig:tmme-image} in the saliency maps with no clear localized region of interest. The model can also generalize well for skin diseases with a high intra-class variance like SCC. The saliency maps from Fig.~\ref{fig:tmme-image} show the ability of the model to localize and segment the region of interest even in low-contrast situations or when hair is present in the image.

The relevancy coefficients in Fig.~\ref{fig:tmme-metadata} highlight clinically important features that define the different lesion pathologies. An example is SCC, which is often found on the face and upper arm/forearm. For MELs, the trained model correctly learned that on average the changing behavior of the lesion is relevant to the diagnosis, while not showing it as criterion for the rest of the pathologies. Additional pathology features intrinsically learned by the model are also highlighted: the role \textit{gender} and \textit{anatomical location} of the lesion play in diagnosing BCC, \textit{age} playing a role in defining a NEV pathology, or the \textit{growth} behavior of SCCs.

Inexperienced medical personnel can rely on such relevancy maps in cases where high inter-class similarity and limited patient metadata makes the visual assessment of the lesion prone to human errors. The model provides additional information via relevancy scores about its internal disease representation helping a dermatologist with a \enquote{learned} definition of the disease.

\subsection{Limitations}
While assessing the risks for the patient, it is not possible to check the quality of the annotations in the dataset used for training and evaluation. This could have important implications for the model's performance. Thus, the model should be validated with other datasets. Moreover, when the algorithm is used outside its intended purpose, it can have a negative impact on patient outcomes due to incorrect classification. Applying the model to out-of-distribution data might lead to overtreatment as currently there is no \enquote{None of the above} class for such cases in the architecture \cite{kneeood}.
In the future, the model should be optimized to maximize the confidence of the classification and an additional class output should be also taken in consideration to represent those inputs who do not cut the confidence threshold of the original disease classes to improve out-of-distribution detection.

\section{Conclusion} \label{section:conclusion}
We developed an interpretable clinical decision support system for skin lesion classification by merging multimodal data sources obtained during the anamnesis of the patient. The proposed method was validated in collaboration with our medical partners to guarantee that both technical and medical aspects of the trained network behavior were adequately covered. The transformer-based architecture demonstrated a robust model for multi-modal data fusion with good results, enabling interpretability of the results to help medical experts comprehend and gain trust in AI-guided decision support systems. 

\review{We believe that the key to the deployment and acceptance of DL-based CDS systems is an interdisciplinary approach, as well as incorporating AI into medical studies to make it more understandable and critically questionable. These approaches to interpretability make a CDS system more user-friendly, allow for better time management, support acceptance, and, in the end, reduce the time between diagnosis and therapy for a better patient care.}

\section*{Acknowledgments}
The research for this article received funding from the German federal ministry of health's program for digital innovations for the improvement of patient-centered care in healthcare (grant agreement no. 2520DAT920).

\renewcommand*{\bibfont}{\footnotesize}
\printbibliography

\end{document}